\documentclass[12pt]{article}

\usepackage{epsfig}
\usepackage{graphics}

\newcommand{\app} [1] {Appendix~\ref{#1}}
\newcommand{\eq} [1] {Eq.(\ref{#1})}
\newcommand{\fig} [1] {Fig.~\ref{#1}}

\newcommand{\tab} [1] {Table~\ref{#1}}

\def\be{\begin{equation}}
\def\ee{\end{equation}}

\begin{document}

\title{
\begin{flushright}
\small DESY 02--166 \\
\small ZU-TH 12/02 
\end{flushright} 
The Reconstruction of Supersymmetric Theories \\ at High Energy Scales}
\author{G.A.~Blair$^{a,b}$, W.~Porod$^{c}$, and P.M.~Zerwas$^{a}$\\[0.5cm] 
\small
$^a$ Deutsches Elektron--Synchrotron DESY, D-22603 Hamburg, Germany \\ \small
$^b$ Royal Holloway, University of London, London,
 UK\\ \small
$^c$ Inst.~f\"ur Theor. Physik, Universit\"at Z\"urich, CH-8057 Z\"urich,
      Switzerland \small
}

\maketitle
\begin{abstract}
The reconstruction of fundamental parameters in supersymmetric theories
requires the evolution to high scales, where the characteristic
regularities in mechanisms of supersymmetry breaking become manifest.
We have studied a set of representative examples in this context: 
minimal supergravity and
a left--right symmetric extension; gauge mediated supersymmetry breaking;
and superstring effective field theories.
Through the evolution of the parameters from the electroweak scale 
the regularities  in different scenarios at the high scales can be
unravelled if 
precision analyses of the supersymmetric particle  sector at
$e^+ e^-$ linear colliders are combined with analyses at the LHC.
\end{abstract}

\section{Introduction}

Extending the Standard Model to a supersymmetric theory 
\cite{Wess:tw,Nilles:1983ge} is an
attractive step which has provided the qualitative understanding of a
diverse set of phenomena in particle physics. Supersymmetry stabilizes
the gap between the Grand Unification scale / Planck scale and the
electroweak scale \cite{Witten:nf}. It allows the unification of the three
gauge couplings at a scale $M_U \simeq 2 \cdot 10^{16}$~GeV in a
straight forward way \cite{GUT}. Radiative electroweak symmetry breaking
relates to the high value of the top mass \cite{Ibanez:fr}. Moreover,
the cold dark matter component in the universe can be identified with
the lightest supersymmetric particle \cite{Ellis:1983ew}.  Above all, local
supersymmetry, requiring the existence of massless spin 2 fields,
provides a rationale for gravity \cite{Freedman:1976xh}.

Supersymmetry is not an exact symmetry in Nature. Unravelling the
breaking mechanism is therefore one of the central issues with this new
concept.  A variety of mechanisms have been proposed, based on rather
different physical ideas. Among these schemes are supergravity
theories \cite{sugra} which have provided the framework for many
phenomenological analyses. The suppression of flavour--changing
neutral reactions is achieved in an automatic form within gauge
mediated supersymmetry breaking \cite{gmsb}. Supersymmetry is broken
in these scenarios in a hidden sector at high and intermediate scales,
respectively, and the breaking is mediated by gravity or gauge
interactions to the visible sector. The breaking, however, may not be
communicated by direct action from the hidden to the visible sector.
This is realized in anomaly mediated supersymmetry breaking models
\cite{Giudice:1998xp} in which supersymmetric particle masses are a
consequence of the superconformal anomaly. In gaugino meditated
supersymmetry breaking \cite{Kaplan:1999ac}, 
supersymmetry is broken on a 3-brane
separated from the 3-brane of the visible sector, and the breaking is
communicated by gauge and Higgs superfields propagating through the
5-dimensional bulk. While in many models of supersymmetry breaking the
gaugino masses are assumed to be universal at the unification scale,
superstring motivated models, in which the breaking is moduli dominated,
as opposed to dilaton dominated scenarios, give rise to
non-universal boundary conditions at the high scale for the gauginos
as well as the sfermion mass parameters \cite{cvetic,Binetruy:2001md}. 
They can be exploited
to determine the parameters of the string effective field theories.

In this report we elaborate on earlier investigations of
Ref.~\cite{Blair:2000gy} in which elements of gravity and gauge
mediated supersymmetry breaking have been considered in realistic
experimental environments of the proton collider LHC \cite{LHC} and
prospective TeV $e^+ e^-$ linear colliders \cite{LC,Aguilar-Saavedra:2001rg}. 
We extend
these investigations in several directions in the present report.

In supergravity inspired models we adopt a scenario close to the 
Snowmass Point SPS\#1 \cite{Allanach:2002nj}. 
In a second step, the previous analysis, based on the Minimal 
Supersymmetric Standard Model, is extended to a left--right
supersymmetric $SO(10)$ model \cite{Zhao:1982rq}. 
The $SO(10)$ symmetry is assumed to be realized
at a scale between the standard $SU(5)$ scale $M_U \simeq 2 \cdot 10^{16}$,
derived from the unification of the gauge couplings, and the Planck scale
$M_P \simeq 10^{19}$~GeV.  The right--handed neutrinos are assumed heavy, with 
masses at intermediate scales
 between $O( 10^{10})$~GeV and $O(10^{15})$~GeV, so that the observed
light neutrino masses are generated by the see-saw mechanism in a natural way
\cite{Gell}.
A rough
estimate of the intermediate scale follows from the evolution of the
mass parameters to the low experimental scale if universality holds at the 
Grand Unification scale.

In the gauge mediated symmetry breaking scenario, the fundamental
scale is expected to be in the range from O(10 TeV) to O($10^6$ TeV).
We present an update and an extension of the earlier analysis. In particular,
the effective
supersymmetry breaking scales, the messenger and supersymmetric mass scales,
 can be reconstructed at the point where
the masses of the sparticles carrying the
same quantum numbers become identical, the characteristic regularity of
gauge mediated supersymmetry breaking.

The anomaly mediated as well as the gaugino mediated SUSY breaking
are technically equivalent to the mSUGRA case and will therefore not
be treated explicitly again.

Among the most exciting schemes rank superstring induced scenarios
(see e.g~\cite{Witten:1985xb,cvetic,Binetruy:2001md} and references therein). 
In this report a string effective field theory,
based on orbifold compactification of the heterotic string, 
will be analyzed in which,
though dominated by the vacuum expectation values of the dilaton field, 
supersymmetry breaking is also affected by the moduli fields.
Such a 
mechanism gives rise to gaugino mass parameters with small but noticeable
departure from universality, and non-universal sfermion mass parameters.
From these mass parameters the  fundamental parameters of the 
string effective field theory, such as
the vacuum expectation values  of the dilaton and the moduli fields,
the moduli/dilaton mixing angle as well as the modular weights
can be derived. In this way high--precision experiments
can provide access to elements which are directly induced by superstrings
\cite{Witten01}.

Extrapolations over many orders of magnitude from the electroweak scale
to scales near the
Planck scale require high--precision measurements at the laboratory
scale \cite{Kane}. Such extrapolations can be performed in practice as
demonstrated in the analysis of the electroweak and strong couplings
at LEP and elsewhere \cite{GUT}. The unification of these couplings
provides the most compelling argument, derived from experiment, in
support of supersymmetry. An initial set of precision data on
supersymmetric particles is expected from LHC experiments if
favourable cascade decays can be exploited to measure mass differences
very precisely \cite{LHC}.  A globally comprehensive
high--precision analysis can only be performed at lepton colliders
\cite{Nojiri96,blair2,LC,Aguilar-Saavedra:2001rg,CLIC}. 
They are expected to be realized in a first phase up
to an energy of about 1 TeV, and in a subsequent second phase up to about
5 TeV. $e^+ e^-$ linear collider designs for the first phase are
being worked out for JLC, NLC and TESLA while the second phase may
be realized in the CLIC technology. TESLA, in particular, can be operated at
very high luminosity. A large number of threshold scans can therefore be
performed which allows model independent
high--precision measurements of the masses of supersymmetric
particles. Chargino, neutralino and slepton masses are expected to be
measured with accuracies at the per--mille level. Very heavy squarks
and gluinos, on the other hand, may be analyzed in detail at CLIC
after their discovery and first analysis
at LHC. However, the  accuracy is presumably
reduced to the per--cent level as a consequence of the decreasing production
cross sections, the non--zero widths of the heavy
particles and the increasing energy smearing due to beam--strahlung.

Starting with observed numbers at the electroweak scale, the
bottom--up approach exhausts all experimental information to the
maximal extent possible in the empirical reconstruction of 
the underlying supersymmetric theory at the high scale.
Finally, the parameters of the fundamental high-scale theory will
become accessible in this way. 
This exploration of GUT and Planck scale physics by
combining high precision with high energy in experiments at hadron and
lepton colliders, is complemented by only a very small
number of other methods, 
{\it notabene} proton decay, likely
neutrino physics, textures of mass matrices, and cosmology. 
In all these individual approaches only
scarce information on the underlying physical theoretical structures 
at the GUT/Planck scale can be
extracted. Any of these methods should therefore be exploited in the maximal
form in order to shed light on the boundaries of the physics area
where  gravity may affect properties and interactions of
particles observed in the laboratory at the electroweak scale. In this way
consequences of 
incorporating the fourth of the fundamental forces into the
particle system could become accessible at laboratory experiments. 

\section{Gravity Mediated SUSY Breaking}

\subsection{Minimal Supergravity -- mSUGRA}

Supersymmetry cannot be broken spontaneously in our eigen--world without 
risking conflict with experimental results. The 
Ferrara--Girardello--Palumbo  mass sum rule \cite{Ferrara:wa}
requires supersymmetric particle masses
below the corresponding Standard Model particle masses in this case
--- in obvious
disagreement with observations.  The elegant concept of spontaneous
symmetry breaking, by non--perturbative gluino condensation for
instance, can be realized, however, in a hidden sector which interacts with our
eigen--world only by gravity. Gravitational interactions generate the
soft supersymmetry breaking terms near the Grand Unification scale / Planck 
scale.  Not compulsory but suggestive, the soft terms may be
universal, {\it i.e.} the gaugino mass parameters and the
scalar mass parameters\footnote {Universality may be broken GUT-scale
threshold corrections, see {\it e.g.} Ref.\cite{HM}. The bottom-up approach 
should enable us to explore this domain in a systematic way 
to the maximum extent possible.}. Being flavour blind, the
suppression of flavour--changing neutral processes can be
realized in a natural way. Moreover, for a heavy top mass 
$m_t \simeq 174$~GeV the breaking of the electroweak symmetry 
$SU(2)_L \times U(1)_Y \to U(1)_{EM}$ can be generated radiatively. 
While at the universality
scale all scalar masses squared are positive, the Higgs mass parameter
$M_{H_2}^2$ turns negative at a scale of about $10^4$~TeV.  
This induces
spontaneous electroweak symmetry breaking at the electroweak scale
where the sum $M_{H_2}^2+|\mu|^2$ becomes negative, leaving however the
strong and electromagnetic gauge symmetries $SU(3)_C$ and
$U(1)_{EM}$ unbroken.

The minimal supergravity scenario mSUGRA is characterized by the universal
parameters 
\begin{center}
\begin{tabular}{lcc}
gaugino mass parameter &:& $M_{1/2}$ \\
 scalar mass parameter &:& $M_0$ \\
 trilinear coupling    &:& $A_0$ 
\end{tabular}
\end{center}
complemented by the phase of $\mu$, the modulus $|\mu|$ determined
by radiative symmetry breaking,  and the mixing angle $\tan \beta$ in the
Higgs sector.
The mass parameters $M_{1/2}$, $M_0$ and the trilinear coupling $A_0$
are defined to be universal at the Grand Unification scale $M_U$. The
unified gauge coupling is denoted by $\alpha_U$ at $M_U$, for the sake 
of simplicity taken real. These
parameters at the GUT scale are related to the low energy parameters
by the supersymmetric renormalization group equations \cite{RGE1,RGE2}
which to leading order generate the evolution for
\begin{center}
\begin{tabular}{lclr}
 gauge couplings &:&  $\alpha_i = Z_i \, \alpha_U$ & (1) \\
  gaugino mass parameters &:& $M_i = Z_i \, M_{1/2}$ & (2) \\
 scalar mass parameters &:&   $M^2_{\tilde j} = M^2_0 + c_j M^2_{1/2} +
        \sum_{\beta=1}^2 c'_{j \beta} \Delta M^2_\beta$  & (3) \\
  trilinear  couplings &:&  $A_k = d_k A_0   + d'_k M_{1/2}$ & (4) 
\end{tabular}
\end{center}
\refstepcounter{equation}
\refstepcounter{equation}
\label{eq:gaugino}
\refstepcounter{equation}
\label{eq:squark} 
\refstepcounter{equation}
The index $i$ runs over the gauge groups $i=SU(3)$, $SU(2)$, $U(1)$.
To leading order, the gauge couplings, and the gaugino and scalar mass
parameters of soft--supersymmetry breaking depend on the $Z$ transporters
\begin{eqnarray}
Z_i = \left[ 1 + b_i {\alpha_U \over 4 \pi}
             \log\left({M_U \over M_Z}\right)^2 \right]^{-1}
\end{eqnarray}
with $b[SU_3, SU_2, U_1] = -3, \, 1, \, 33 / 5$;
the scalar mass parameters depend 
also on the Yukawa couplings $h_t$, $h_b$, $h_\tau$
of the
top quark, bottom quark and $\tau$ lepton.
The coefficients $c_j$ [$j=L_l, E_l, Q_l, U_l, D_l, H_{1,2}$; $l=1,2,3$] 
for the slepton and squark doublets/singlets of generation $l$, 
and for the two Higgs doublets
are linear combinations of the evolution
coefficients $Z_i$; the coefficients $c'_{j \beta}$ are of order unity. 
The shifts $\Delta M^2_\beta$ are nearly zero for the first two families of 
sfermions but they can be rather large for the third family and for the 
Higgs mass
parameters, depending on the coefficients $Z_i$, 
the universal parameters $M^2_0$, $M_{1/2}$ and $A_0$,
and on the Yukawa couplings $h_t$, $h_b$, $h_\tau$.
The coefficients $d_k$ of the trilinear
couplings $A_k$ [$k=t,b,\tau$]  
depend on the corresponding Yukawa couplings 
and they are approximately unity for the
first two generations while being O($10^{-1}$) 
and smaller if the Yukawa couplings are
large; the coefficients $d'_k$, depending on gauge 
and Yukawa couplings, are of order unity.
Beyond the approximate solutions shown explicitly, the evolution equations 
have been solved numerically in the present analysis  to
two--loop order \cite{RGE2} and threshold effects have been
incorporated at the low scale \cite{bagger}.

These parameters enter  the mass--matrices for the various particles. In 
the case
of charginos ${\tilde \chi}^+_m$ [$m=1,2$] the $2 \times 2$ mass matrix 
reads as
\begin{eqnarray}
M_{{\tilde \chi}^+} = \left( \begin{array}{cc}
   M_2 & \sqrt{2} m_W \cos\beta\\
  \sqrt{2}  m_W  \sin\beta& \mu 
  \end{array} \right)
\end{eqnarray}
while the mass matrix for neutralinos ${\tilde \chi}^0_n$ [$n=1,..,4$]
is a $4\times 4$ matrix,
\begin{eqnarray}
M_{{\tilde \chi}^0} = \left( \begin{array}{cccc}
   M_1 & 0 & - m_Z  \cos\beta \sin\theta_W & m_Z  \sin\beta \sin\theta_W \\ 
   0 & M_2 &  m_Z  \cos\beta \cos\theta_W & - m_Z  \sin\beta \cos\theta_W  \\
 - m_Z  \cos\beta \sin\theta_W &  m_Z  \cos\beta \cos\theta_W & 0 & -\mu \\
    m_Z  \sin\beta \sin\theta_W & - m_Z  \sin\beta \cos\theta_W & -\mu&0 \\
  \end{array} \right)
\end{eqnarray}
Exploiting all the information available from a 
linear collider, both mass matrices can be reconstructed even the in
case of complex parameters \cite{Choi:2000ta}. For large values, $\tan\beta$
needs supplementary analyses in the Higgs sector \cite{Gunion:2001qy}.

Assuming that the sfermion generations mix only weakly
the mass matrices of the third generation
sfermions can be written
as:
\begin{eqnarray}
M^2_{\tilde{f}} = \left( \begin{array}{cc}
                        m^2_{\tilde{f}_L} & a_f m_f \\
                        a_f m_f & m^2_{\tilde{f}_R}
                       \end{array}
                 \right)
\label{eq:sqmmat}
\end{eqnarray}
with
\begin{eqnarray}
  &&\hspace{-8mm} m^2_{\tilde f_L} = 
  M^2_{\tilde F_L} + (T^3_f - e_f \sin^2\theta_W)\cos 2\beta \, m_Z^2 
       + m^2_f , 
  \label{eq:msfl}\\
  &&\hspace{-8mm} m^2_{\tilde f_R} = 
  M^2_{\tilde F_R} + e_f \sin^2\theta_W \cos 2\beta \, m_Z^2 + m^2_f , 
  \label{eq:msfr}\\
  &&\hspace{-8mm} a_t \equiv A_t - \mu \cot \beta, \hspace{2mm} 
  a_b \equiv A_b - \mu \tan \beta, \hspace{2mm} 
  a_{\tau} \equiv A_{\tau} - \mu \tan\beta,
  \label{eq:offdiag}
\end{eqnarray}
where $e_f$ and $T^3_f$ are the electric charge and the third component
of the weak isospin of the sfermion $\tilde f$; $M_{\tilde F_L}=M_{\tilde Q}$ 
for ${\tilde f}_L = \tilde t_L,\, \tilde b_L$, $M_{\tilde F}=M_{\tilde L}$ for
$\tilde f_L = \tilde \tau_L, \tilde \nu_\tau$; 
$M_{\tilde F_R} = M_{\tilde U},\, M_{\tilde D},\,M_{\tilde E}$ 
for $\tilde{f}_R = \tilde t_R,\, \tilde b_R,\, \tilde \tau_R$,
respectively;  $m_f$ is the mass of the corresponding fermion.
Also in this case it has been shown that the mass matrix can be reconstructed
\cite{StauNew,Bartl:1997yi}. The mass matrices for the first two generation
sfermions have the same structure. However, due to the small fermion masses
the mixing between the $L/R$ sfermions can be neglected in general.
In the fits for the parameters we have used the complete one--loop
mass matrices as given in \cite{bagger}. For the Higgs bosons
also the two--loop contributions \cite{Degrassi:2001yf} are included.

The mSUGRA point we have analyzed in detail, was chosen close to 
the Snowmass Point SPS\#1
\cite{Allanach:2002nj}, except for the scalar mass parameter
$M_0$ which was taken slightly larger for merely
illustrative purpose: $M_{1/2} = 250$~GeV, $M_0 = 200$~GeV, $A_0 = -100$~GeV,
$\tan \beta = 10$ and $sign(\mu) = +$. 
The initial ``experimental'' values, 
have been generated by evolving the universal parameters down to the 
electroweak scale according to standard
procedures \cite{aarason,bagger}.

The parameters chosen are compatible with the present results of 
low--energy experiments which they affect by virtual contributions, and they
are also compatible with dark--matter estimates \cite{Primack:2002th}: 
BR($b\to s \gamma) = 2.7 \cdot 10^{-4}$, 
$\Delta[g - 2]_\mu = 17 \cdot 10^{-10}$, $\Delta \rho = 38 \cdot 10^{-5}$
and $\Omega h^2 = 0.4$. We have used the formulas given in \cite{CLEO}
for the computation of $b\to s \gamma$, those given in \cite{Drees90}
for $\Delta \rho$ and those given in \cite{Ibrahim:1999hh} for 
$\Delta[g - 2]_\mu$. $\Omega h^2$ has been calculated using the program
of Ref.\cite{ds}.

The five basic parameters define the experimental observables, including
supersymmetric particle masses and production cross sections. They are
endowed with errors as expected for 
threshold scans as well as measurements  
in the continuum  at $e^+ e^-$ linear colliders (LC). 
Major parts of the LC
analysis can be performed for energies below 1 TeV, some of the squarks
require energies  above 1 TeV. Estimates are based on integrated 
LC luminosities of  1 ab$^{-1}$.
The errors given in  Ref.\cite{blair2} are scaled
in proportion to the masses of the spectrum. 
Typical examples are shown in Table~\ref{tab:masserrors}.
The LC errors on the squark masses, see {\it e.g.} Ref.\cite{Feng},
are set to an average value of 10 GeV [similar errors may also be
obtained if the precisely measured mass differences at the LHC
are combined with high-precision measurements of the low-lying states
at the LC];
varying this error within a factor two does not change the conclusions
significantly since the measurement of the cross sections
provides the maximal sensitivity in this sector.
For the cross-sections we use purely statistical errors, while
assuming a (conservative) reconstruction efficiency of 20\%.  
In addition
the mass errors on the lightest gauginos were inflated with
respect to earlier analyses to be conservative in advance of
detailed experimental analyses of models with higher values of
$\tan\beta$.
[Parameter combinations from the 
fits to the spectrum and the cross sections which lead 
to charge and/or color breaking minima \cite{Casas},
are not accepted.]
These observables are interpreted as the experimental input values for
the evolution of the mass parameters in the bottom-up
approach to the Grand Unification scale. 
\begin{table}
\caption[]{\it Representative experimental mass errors used in the
fits to the mass spectra; with the exception of the gluino mass, all
the other parameters are based on LC measurements.}
\label{tab:masserrors}
\begin{center}
\begin{tabular}{c|cc||c|cc}
Particle           & M(GeV) & $\Delta$ M(GeV) &
Particle           & M(GeV) & $\Delta$ M(GeV)\\\hline \hline
$h^0$              &  113.33      & 0.05  &
$\tilde{\nu}_{e L}$    & 256.79    &       0.11   \\
$H^0$ & 436.1 & 1.5 & $\tilde{e}_L$      & 269.1 & 0.3 \\
$A^0$              &  435.5        &  1.5  &
 $\tilde{e}_R$      & 224.82    &       0.15\\
$H^+$ & 443.3 & 1.5 & $\tilde{\nu}_{\tau L}$ & 255.63    &       0.95   \\
$\tilde{\chi}^+_1$         & 183.05         & 0.15   &
$\tilde{\tau}_1$   & 217.7    &       1.00 \\
$\tilde{\chi}^+_2$         & 383.3         & 0.3   &
$\tilde{\tau}_2$   & 271.5    &       0.9 \\
$\tilde{\chi}^0_1$         & 97.86        &  0.2 &
$\tilde{u}_L$      & 589    &       10  \\
$\tilde{\chi}^0_2$         & 184.6        &  0.3 &
$\tilde{u}_R$      & 572    &       10  \\
$\tilde{\chi}^0_3$         & 365.5        &  0.3 &
$\tilde{d}_R$      & 572    &       10  \\
$\tilde{\chi}^0_4$         & 383.0        &  0.7 &
$\tilde{t}_1$      & 412    &       10  \\
$\tilde{g}$        & 598    &   10   &
$\tilde{t}_2$      & 600    &       10  \\ \hline
\end{tabular}
\end{center}
\end{table}

\subsubsection{Gauge Couplings}

The presumably strongest support, though indirect, 
for supersymmetry is related to the
tremendous success of this theory in predicting the unification of the
gauge couplings \cite{GUT}. The precision, being  at the per--cent level, is
surprisingly high after extrapolations over
fourteen orders of magnitude in the energy 
from the electroweak scale to the unification scale $M_U$. 
Conversely, the
electroweak mixing angle has been predicted in this approach at the
per--mille level. The evolution of the gauge couplings from 
low energy  to the GUT scale $M_U$ is carried out in the two--loop
accuracy. 
The gauge couplings $g_1$,
$g_2$, $g_3$ and the Yukawa couplings are calculated
in the $\overline{DR}$ scheme
by adopting the shifts given in \cite{bagger}.
These parameters are evolved to $M_U$ using 2--loop RGEs \cite{RGE2}. 
At 2-loop order the gauge couplings do not meet
exactly \cite{Weinberg:1980wa,Hall:1980kf}, the
differences owing to threshold effects at the unification
scale $M_U$ which leave us with an ambiguity in the definition of
$M_U$. In this report we define $M_U$ as the scale, {\it ad libitum}, 
where $g_1 = g_2$ in the RGE evolution.
The non--zero 
difference $g_1 - g_3$ at this scale is then attributed to threshold effects
of particles with masses of order $M_U$. The quantitative evolution implies 
important constraints on the particle content at $M_U$
\cite{Ross:1992tz}-\cite{MurPie}.

\begin{figure*}
\setlength{\unitlength}{1mm}
\begin{center}
\begin{picture}(160,85)
\put(-14,-97){\mbox{\epsfig{figure=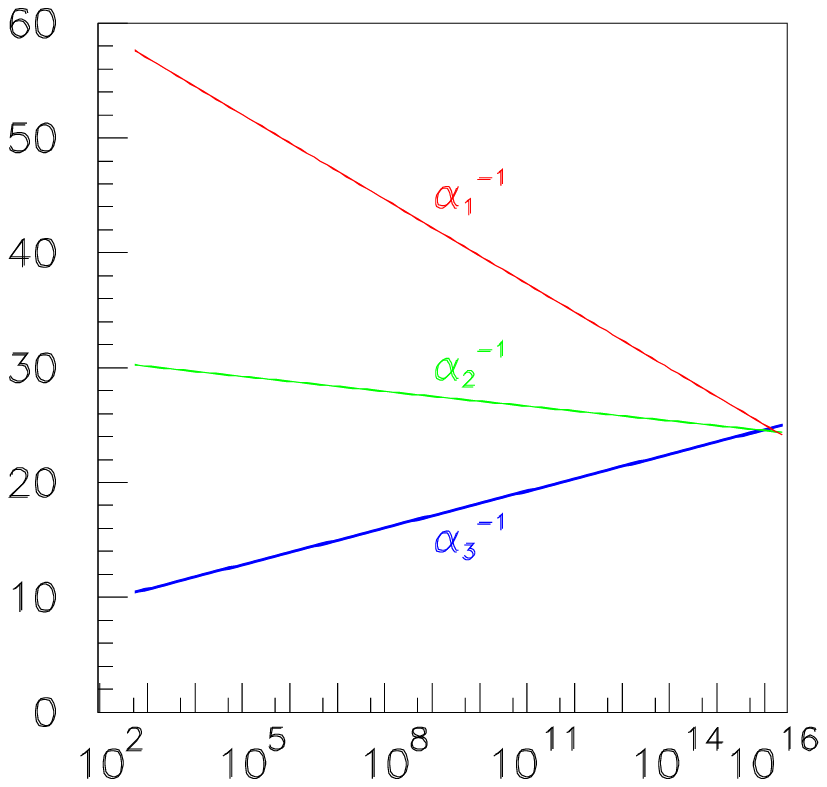,
                                   height=19.cm,width=19.5cm}}}
\put(84,-3){\mbox{\epsfig{figure=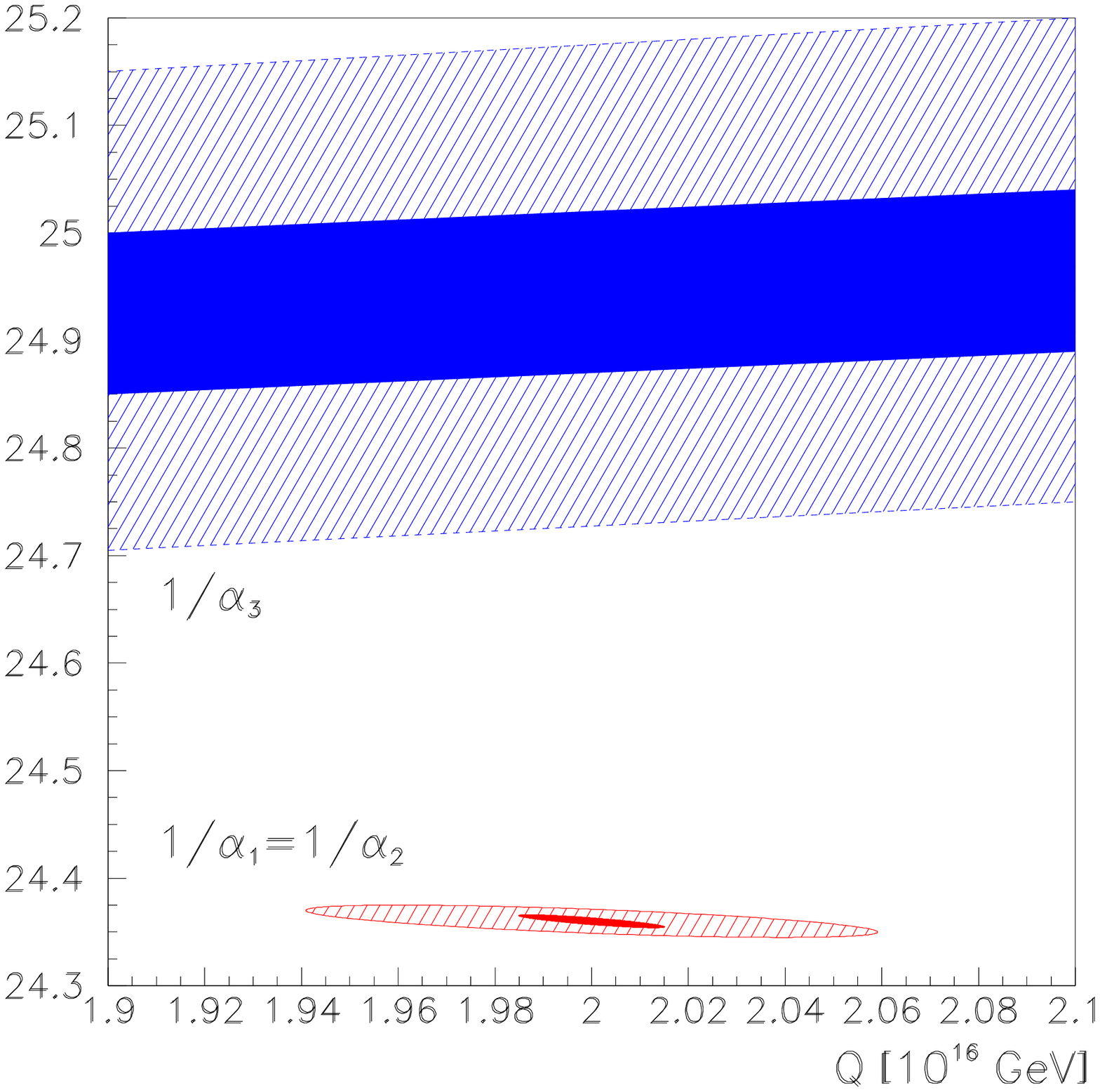,
                                   height=8.5cm,width=8.5cm}}}
\put(-3,80){\mbox{\bf a)}}
\put(84,80){\mbox{\bf b)}}
\put(60,-3){\mbox{Q~[GeV]}}
\end{picture}
\end{center}
\caption{\it a) Running of the inverse gauge couplings. b) Determination 
             of $M_U$, $\alpha_U^{-1}$; the unification point $U$ is
             defined by the meeting point of $\alpha_1$ with $\alpha_2$.
             The wide error bands are based on present data, the narrow
             bands demonstrate the improvement expected by future GigaZ 
             analyses.}
\label{fig:gauge}
\end{figure*}

Based on the set of low--energy gauge and Yukawa parameters 
$\{\alpha(m_Z)$, $\sin^2
\theta_W$, $\alpha_s(m_Z)$, $Y_t(m_Z)$, $Y_b(m_Z)$, $Y_\tau(m_Z)\}$
the evolution of the inverse couplings $\alpha_i^{-1}$ $[i=U(1)$, $SU(2)$,
$SU(3)]$ is depicted in \fig{fig:gauge}a. The evolution is performed for
the mSUGRA reference point defined above. Unlike earlier analyses, the
low--energy thresholds of supersymmetric particles can be calculated
in this framework
exactly without reference to effective SUSY scales. The broken error
ellipse in \fig{fig:gauge}b, derived for [$M_U, \alpha_U$] by
requiring $g_1=g_2$,
corresponds to the present experimental accuracy of the gauge
couplings \cite{PDG}: 
$\Delta \{\alpha^{-1}(m_Z)$, $\sin^2\theta_W$, $\alpha_s(m_Z)\}$ 
$=\{ 0.03, 1.7 \cdot 10^{-4}, 3 \cdot 10^{-3} \}$. 
The full ellipse demonstrates the improvement for the absolute errors
$\{ 10^{-3}, 10^{-5},10^{-3}  \}$ after operating GigaZ 
\cite{Monig:2001hy,Erler:2000jg}.  
The expected accuracies in $M_U$ and $\alpha_U$
are summarized in the values given in \tab{tab:gauge}.
The difference between the unification point in the ellipse and the value of
$g_3$ is accounted for by contributions from high scale physics, 
colour-triplet Higgs fields, for example.
Thus, for a typical set of SUSY parameters, the evolution of the gauge
couplings from low  to high scales leads to a precision of 1.5 per--cent
of the Grand Unification picture.

\begin{table}[th]
\caption{\it Expected errors on $M_U$ and $\alpha_U$ for the mSUGRA reference
 point, derived for the present level of accuracy and compared with GigaZ.}
\label{tab:gauge}
\begin{center}
\begin{tabular}{c||c|c}
 & Now & GigaZ \\
\hline \hline
$M_U$ & $(2.00 \pm 0.06)\cdot 10^{16} \, \rm {GeV}$ & 
           $ (2.000 \pm 0.016) \cdot 10^{16} \, \rm {GeV}$ \\
$\alpha_U^{-1}$ & $  24.364 \pm 0.015 $ &  $ 24.361 \pm 0.007 $\\ \hline
\end{tabular}
\end{center}
\end{table}

\subsubsection{Gaugino and Scalar Mass Parameters}

\begin{figure*}
\setlength{\unitlength}{1mm}
\begin{center}
\begin{picture}(160,140)
\put(-4,0){\mbox{\epsfig{figure=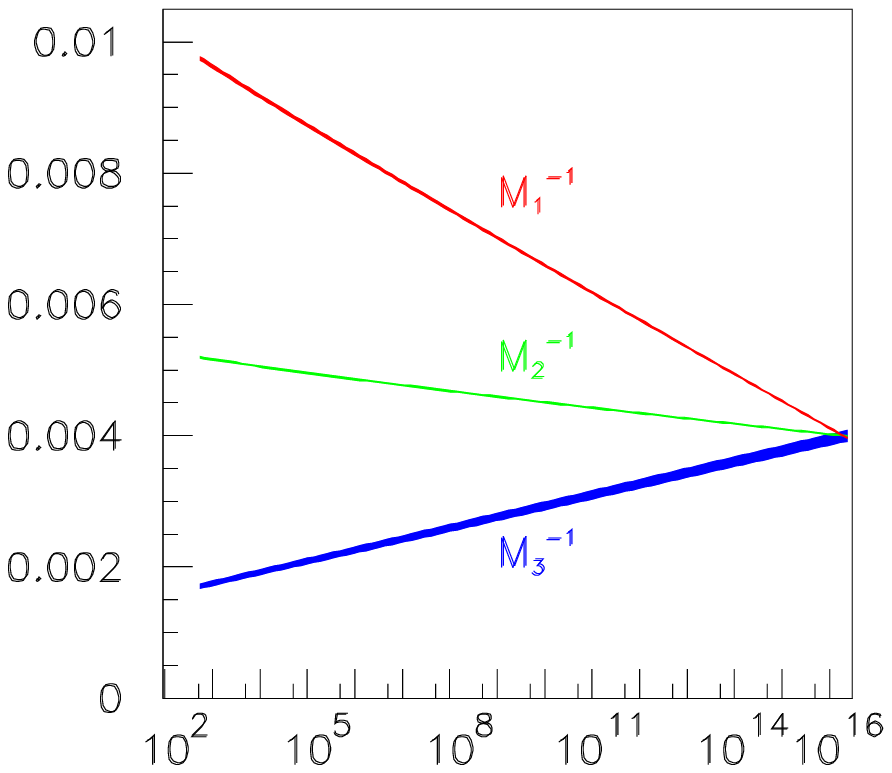,height=17cm,width=18cm}}}
\put(83,81){\mbox{\epsfig{figure=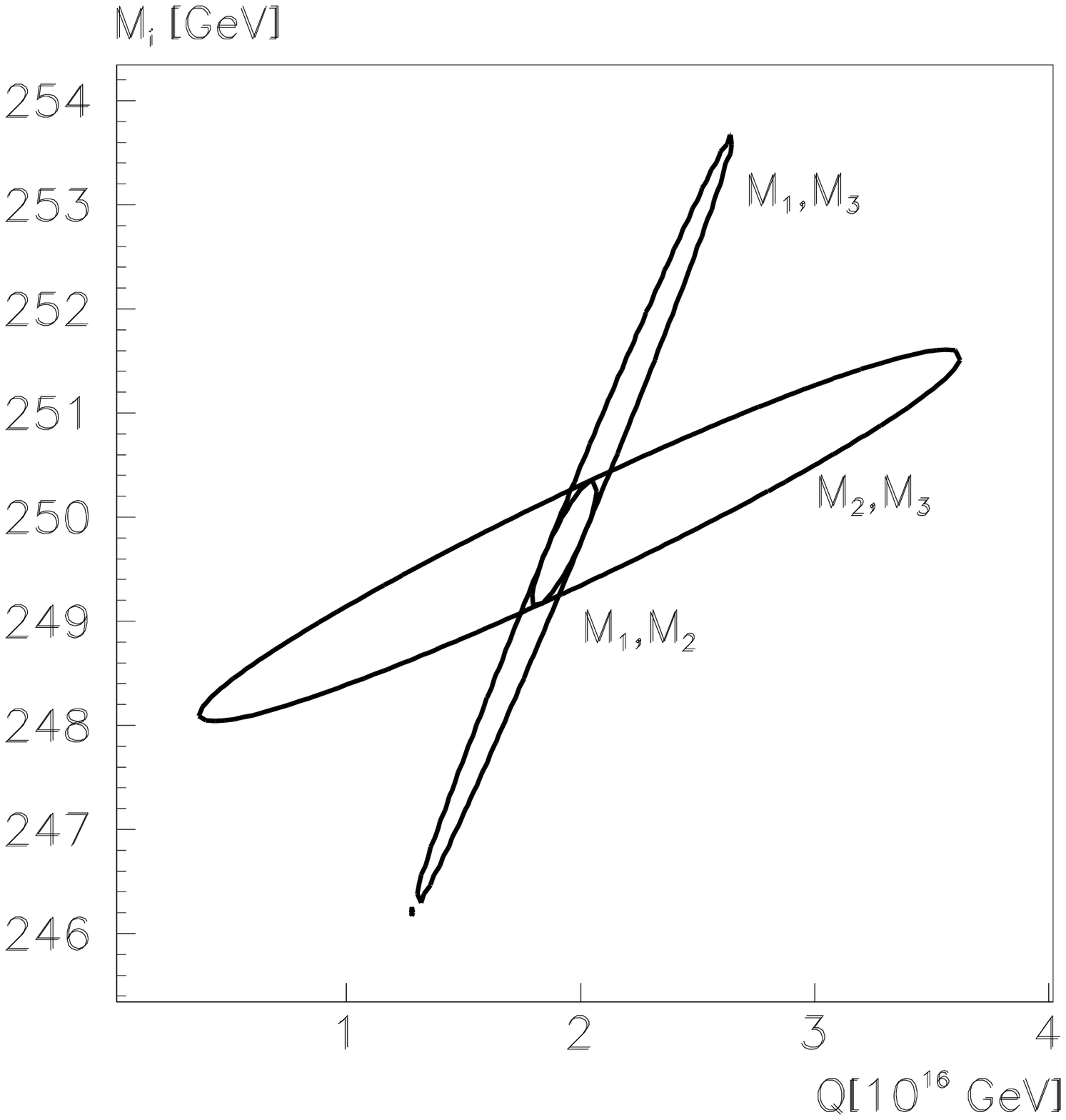,
                                   height=8.0cm,width=8.5cm}}}
\put(-4,-86){\mbox{\epsfig{figure=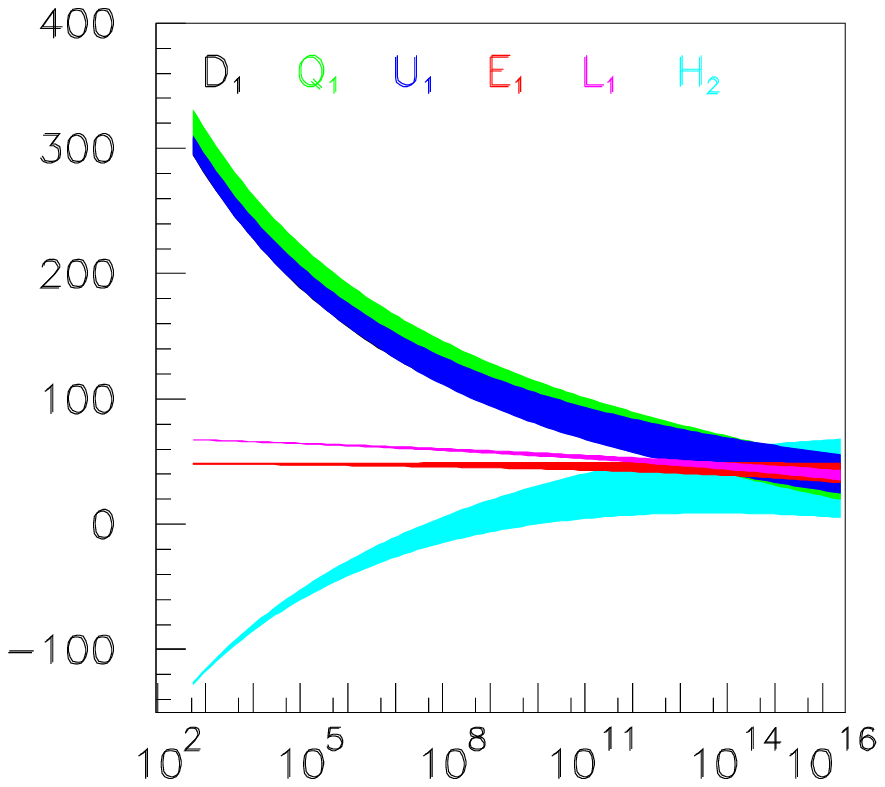,height=17cm,width=18cm}}}
\put(78,-86){\mbox{\epsfig{figure=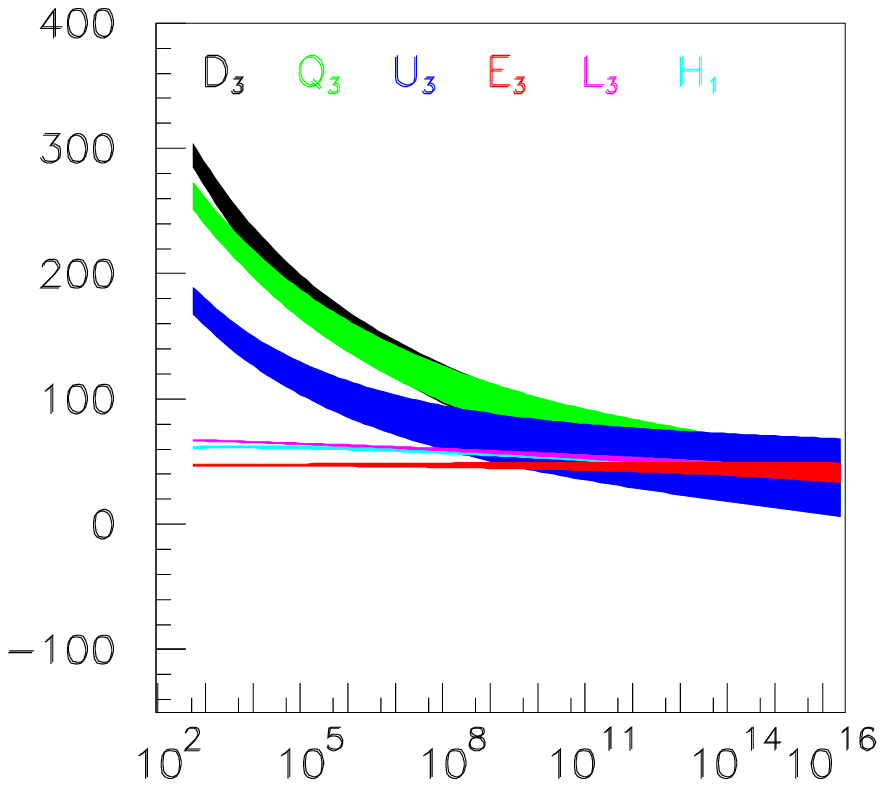,height=17cm,width=18cm}}}
\put(-1,158){\mbox{\bf (a)}}
\put(14,156){\mbox{$1/M_i$~[GeV$^{-1}$]}}
\put(65,82){\mbox{$Q$~[GeV]}}
\put(80,159){\mbox{\bf (b)}}
\put(-1,73){\mbox{\bf (c)}}
\put(14,71){\mbox{$M^2_{\tilde j}$~[$10^3$ GeV$^2$]}}
\put(65,-3){\mbox{$Q$~[GeV]}}
\put(80,73){\mbox{\bf (d)}}
\put(95,71){\mbox{$M^2_{\tilde j}$~[$10^3$ GeV$^2$]}}
\put(147,-3){\mbox{$Q$~[GeV]}}
\end{picture}
\end{center}
\caption{{\bf mSUGRA:} {\it  Evolution, from low to high scales, of 
(a) gaugino mass parameters, and (b) unification of gaugino mass parameter
   pairs; 
 (c) evolution of first--generation sfermion mass parameters and 
     the Higgs mass parameter $M^2_{H_2}$;
 (d) evolution of third--generation sfermion mass parameters and  
     the Higgs mass parameter $M^2_{H_1}$. 
 The mSUGRA point probed is defined by the
parameters $M_0 = 200$~GeV, 
$M_{1/2} = 250$~GeV, $A_0$ = -100~GeV, $\tan \beta = 10$, 
and $\mathrm{sign}(\mu) = (+)$.
[The widths of the bands indicate the 1$\sigma$ CL.]
}}
\label{fig:sugra}
\end{figure*}

The results for the evolution of the mass parameters to the GUT scale $M_U$
are shown in \fig{fig:sugra}.
 \fig{fig:sugra}(a)  presents the evolution of the gaugino
parameters $M_i^{-1}$ which clearly is under excellent control, 
as is the extrapolation
of the slepton mass parameter squared of the first (and second) and the
third generation
in Fig.~\ref{fig:sugra}(c) and (d), respectively. The accuracy
deteriorates for the squark mass parameters and for the Higgs mass parameter
$M^2_{H_2}$.
The origin of the differences between the errors for slepton, squark and
Higgs mass parameters can be traced back to the numerical 
size of the coefficients
in Eqs.~(\ref{eq:squark}). Typical examples using the formulas presented
in Appendix~\ref{sec:AppRGE} evaluated at $Q=500$~GeV read as
follows:
\begin{eqnarray}
M^2_{\tilde L_{1}} &\simeq& M_0^{2} + 0.47 M^2_{1/2} \\
M^2_{\tilde Q_{1}} &\simeq& M_0^{2} + 5.0 M^2_{1/2}  \\
M^2_{\tilde H_2} &\simeq&  -0.03 M_0^{2} - 1.34 M^2_{1/2}
           + 1.5 A_0 M_{1/2} + 0.6 A^2_0 \\
|\mu|^2 &\simeq& 0.03 M_0^{2} + 1.17 M^2_{1/2}
           - 2.0 A_0 M_{1/2} - 0.9 A^2_0
\end{eqnarray}
While the coefficients for sleptons are of order unity, the coefficient $c_j$
for squarks grows very large,  $c_j \simeq 5.0$, so that small errors
in $M^2_{1/2}$ are magnified by nearly an order of magnitude in the solution
for $M_0$. By close inspection of Eq.(\ref{eq:squark}) for the Higgs mass
parameter it turns out that 
the formally leading 
$M^2_0$ part is nearly cancelled by the $M^2_0$ part
of $c'_{j,\beta} \Delta M_\beta^2$. Inverting Eq.(\ref{eq:squark}) for
$M^2_0$ therefore gives rise to large errors in the Higgs case.
A representative
set of mass values and the associated errors,  
after evolution from the electroweak scale to $M_U$, is
presented in Table~\ref{tab:parvalues_a}.
The corresponding error ellipses for the unification of the 
gaugino masses are shown in
\fig{fig:sugra}(b).

Extracting the trilinear parameters $A_k$ is difficult and
more refined analyses
based on sfermion cross sections and Higgs and/or sfermion decays are
necessary to determine these parameters accurately.
$A_t$ can be obtained from the mixing angle of the stop sector
by measuring the stop production cross section in $e^+ e^-$
annihilation with different electron and/or positron polarizations
\cite{Bartl:1997yi}. In the cases $A_b$ and $A_\tau$ the situation is
more difficult, because these parameters influence the mixing angle
in the sbottom and stau sector only weakly as evident from \eq{eq:offdiag}.
In these cases the $\tilde b$ and $\tilde \tau$ couplings to the Higgs bosons
must be measured,
because these couplings include terms directly proportional to $A_k \tan\beta$.
For instance, by analysing the decays 
$\tilde \tau_2 \to A^0 \tilde \tau_1$, $h^0 \tilde \tau_1$ and
$H^0 \tilde \tau_1$, $A_\tau$ can be extracted within 10\% \cite{StauNew}. 
If these modes are kinematically forbidden, the couplings can
either be measured in the decays of the heavier Higgs bosons, 
as $H^0, A^0 \to  \tilde \tau_1 \tilde \tau_1$, or by means of the cross
sections for processes such as $e^+ e^- \to \tilde  \tau_1 \tilde \tau_1 h^0$.
Similar procedures are expected to apply for $A_b$. In certain areas of
SUSY parameter space, the trilinear couplings can also be extracted from
measurements of the degree of fermion polarization \cite{Boss} in sfermion 
decays $\tilde t, \tilde b$ and $\tilde \tau$.   

\begin{table}
\caption[]{\it Representative gaugino/scalar mass parameters and couplings
as determined
at the electroweak
scale and evolved to the GUT scale in the mSUGRA scenario;
based on LHC
and LC simulations. $M^2_{L_{1,3}}$, $M^2_{Q_{1,3}}$ are the slepton and
squark isodoublet parameters of
the first and third family whereas $M^2_{E_{1,3}}$, $M^2_{U_{1,3}}$ and
$M^2_{D_{1,3}}$ are the  the slepton and
squark isosinglet parameters of
the first and third family. [The errors quoted correspond to 1$\sigma$.]}
\label{tab:parvalues_a}
\begin{center}
\begin{tabular}{c||c|c}
 &  Exp.~Input &  GUT Value \\ \hline   \hline
 $M_1$~[GeV] & 102.31 $\pm$  0.25 &  $250.00 \pm  0.33$ \\
 $M_2 $~[GeV] &  192.24 $\pm$  0.48      &  $250.00 \pm  0.52$ \\
 $M_3 $~[GeV] & 586  $\pm$  12   &  $250.0    \pm   5.3$  \\ \hline
$\mu$         & 358.23  $\pm$ 0.28     &  $355.6 \pm  1.2    $  \\
\hline
 $M^2_{L_1} $~[GeV$^2$] & $( 6.768  \pm  0.005)\cdot 10^4$
                &  $(3.99  \pm  0.41) \cdot 10^4$  \\
 $M^2_{E_1} $~[GeV$^2$] & $(4.835  \pm  0.007) \cdot 10^4$
  &  $(4.02  \pm  0.82)  \cdot 10^4 $ \\
 $M^2_{Q_1} $~[GeV$^2$] &  $(3.27 \pm  0.08)\cdot 10^5$
               &  $(3.9  \pm  1.5) \cdot 10^4$ \\
 $M^2_{U_1} $~[GeV$^2$] &  $(3.05 \pm  0.11)\cdot 10^5$
               &  $(3.9  \pm  1.9) \cdot 10^4$ \\
 $M^2_{D_1} $~[GeV$^2$] &  $(3.05 \pm  0.11)\cdot 10^5$
               &  $(4.0  \pm  1.9)  \cdot 10^4$
\\ \hline 
 $M^2_{L_3} $~[GeV$^2$] & $(6.711 \pm  0.050)\cdot 10^4$
   &  $(4.00  \pm  0.41)  \cdot 10^4 $  \\
 $M^2_{E_3} $~[GeV$^2$] & $(4.700 \pm  0.087)\cdot 10^4$
  &  $(4.03  \pm  0.83) \cdot 10^4 $  \\
 $M^2_{Q_3} $~[GeV$^2$] &  $(2.65 \pm  0.10) \cdot 10^5$
  &  $(4.1  \pm  3.0)  \cdot 10^4 $ \\
 $M^2_{U_3} $~[GeV$^2$] &  $(1.86 \pm  0.12)\cdot 10^5$
  &  $(4.0  \pm  3.6)  \cdot 10^4 $ \\
 $M^2_{D_3} $~[GeV$^2$] &  $(3.03 \pm  0.12)\cdot 10^5$
  &  $(4.0  \pm  2.6)  \cdot 10^4 $ \\
\hline
 $M^2_{H_1} $~[GeV$^2$] &  $(6.21 \pm  0.08)\cdot 10^4$  &
  $ (4.01  \pm  0.54)  \cdot 10^4 $ \\
 $M^2_{H_2}$~[GeV$^2$] &  $(-1.298 \pm 0.004)\cdot 10^5$ &
  $(4.1  \pm  3.2) \cdot 10^4 $\\
 $A_t $~[GeV] & -446 $\pm$ 14   &  -100 $\pm$ 54   \\ \hline
 $\tan\beta$ & $9.9  \pm  0.9$ & --- \\ \hline
\end{tabular}
\end{center}
\end{table}

\begin{table}
\caption[]{\it Comparison of the ideal parameters with the
experimental expectations for the particular mSUGRA reference
point analyzed in this report. [All mass parameters are given 
in units of $GeV$.]}
\label{tab:parvalues_b}
\begin{center}
\begin{tabular}{c||c|c}
                &  Ideal           & Experimental Error \\ \hline  \hline
$M_U$           & $2\cdot 10^{16}$ &  $1.6  \cdot 10^{14}$       \\
$\alpha_U^{-1}$ &   24.361          &     0.007     \\ \hline
$M_\frac{1}{2}$ & 250              & 0.08     \\
$M_0$           & 200              & 0.09     \\
$A_0$           & -100             & 1.8      \\  \hline
$\mu$           & 358.23           & 0.21     \\
$\tan\beta $    &  10              & 0.1      \\  \hline
\end{tabular}
\end{center}
\end{table}

The unified value $A_0$ of the
$A_t$ coupling, the best measured coupling among the $A_k$ parameters,
is shielded by the pseudo--fixed point behaviour of $A_t$
\cite{Allanach} since $d_t \simeq 0.2$
is small compared to $d'_t \simeq 2$. The impact of the other trilinear 
couplings on physical observables is weak so that 
large experimental errors are expected. As a result, the universal 
character of the fundamental parameter $A_0$ cannot be determined 
as precisely as the other parameters at the GUT scale.  
Although the trilinear couplings $A_b$ and $A_\tau$ have only little impact on
physical observables, they do strongly influence the running of the third
generation sfermion mass parameters as well as the Higgs mass parameters.
The error propagation is stabilized if $A_{\tau}$ and $A_b$ can be
measured in the way outlined above. [Otherwise  
the errors would increase by an order of magnitude.]
The detailed analysis in this report has been based on the 
auxiliary assumption that $A_b$ and $A_\tau$ are within 1$\sigma$ of $A_t
= A_0$ at $M_U$; this assumption is conservative if the envisaged experimental
analyses of $A_{\tau}$ and $A_b$ can be performed at the electroweak scale 
in practice. 

Even though the auxiliary assumption seems conservative, given the
size of the error on $A_0$ determined from $A_t$, dedicated phenomenological
and experimental analyses of the third family must be developed,
as indicated above,
to improve the measurement of the associated parameters, in particular
in view of the evolution of the Higgs mass parameter which induces
electroweak symmetry breaking.

Inspecting Figs.~\ref{fig:sugra}(c) and (d) leads to the conclusion that a
blind top-down approach eventually may generate an incomplete picture. 
Global fits based on mSUGRA without allowing for deviations
from universality, are dominated by $M_{1,2}$ and the slepton mass
parameters due to the pseudo-fixed point behaviour of the squark mass
parameters.  Therefore, the structure of the theory in the squark sector
is not scrutinized stringently at the unification scale
in the top-down approach let alone the Higgs sector. 
By contrast, the bottom-up approach demonstrates very clearly the extent
to which the theory can be tested at the high scale quantitatively.
The quality of the test is apparent from \tab{tab:parvalues_a}, in which 
the evolved gaugino
values should reproduce the universal mass $M_{1/2} = 250$~GeV and all
the scalars the mass $M_0 = 200$~GeV. They are compared with the global
mSUGRA fit of the universal parameters in \tab{tab:parvalues_b}.

\subsection{Left--Right Supergravity}

It is generally accepted that neutrinos are massive particles, though
at a very low scale. Supersymmetric scenarios like MSSM and mSUGRA
must therefore be extended to incorporate the right--handed neutrino
degrees of freedom.  Since the complexity grows strongly with the rising
number of parameters,
it is useful, in a first attempt, to analyze the system in
characteristic scenarios based on compelling  physical
assumptions. In particular, we will assume that the small neutrino
masses are generated by the seesaw mechanism \cite{Gell}. 
Moreover, we will assume
hierarchies for the heavy neutrino masses as well as the neutrino
Yukawa couplings similar to the up--type particles in the quark
sector; such a scheme, suggested by $SO(10)$ GUT, is compatible with the
data collected in low--energy neutrino experiments \cite{Wyler}.

This scenario can be embedded in a Grand Unified $SO(10)$ theory with
the following breaking pattern of the symmetries. The $SO(10)$
symmetry is realized between the Planck scale $M_P$ and a scale
$M_{SO(10)}$ at which the symmetry breaks to $SU(5)$.  The scale
$M_{SO(10)}$ is assumed above the scale $M_U$ where $SU(5)$ breaks to
the symmetry group $SU(3)_C \times SU(2)_L \times U(1)_Y$ of the
Standard Model. At the scale $M_U$ the gauge couplings split and the
effective theory is the MSSM plus right--handed neutrinos with masses
of order $10^{9}$ to $10^{15}$~GeV. Below this mass scale the
right--handed neutrinos freeze out and the MSSM is effectively realized in its
standard form. The relevant SUSY parameters are summarized in
\tab{tab:scales}.  It is less obvious that $M_U$ associated with the
$SU(5)$ symmetry is the scale where the gaugino and scalar mass
parameters are universal.  The supporting argument for this point
is derived empirically from the
unification of the gauge couplings.  Nevertheless, 
the subsequent analysis will be based on this hypothesis which, of course,
is a clear target for confirmation or rejection in the bottom--up
approach we investigate\footnote{Potential sources of deviations from
  this picture can easily be illustrated by assuming $M_{SO(10)}$ as
  the scale proper of universality: The Yukawa interactions contribute
  differently to the running of the $M^2_{\widetilde{10}}$, 
  $M_{\tilde{5}}^2$, $M^2_{\tilde \nu_R}$; 
  the same holds true for the $A$ parameters \cite{Polonsky:1994sr}. Moreover,
  different $D$--term contributions to the scalar masses are in
  general generated by the breaking mechanism from $SO(10)$ to
  $SU(5)$ \cite{Kolda:1995iw}.}.

\begin{table}
\caption{\it Scales and soft SUSY breaking parameters of the effective 
         left--right supergravity theory analyzed in this report.}
\label{tab:scales}       
\begin{center}
\begin{tabular}{l||l|l}
 Scale & Gauge Group & Parameters  \\
\hline\hline
$M_{P}$ -- $M_{SO(10)}$ & $SO(10)$ & $M_{1/2}$, $M^2_{\widetilde{16}}, A_0$ \\
$M_{SO(10)}$ --$M_U$ & $SU(5)$ &
        $M_{1/2}$, $M^2_{\widetilde{10}}$, $M^2_{\tilde{5}}$\\
             &         & $M^2_{\tilde \nu_R}$, $A_{10}$, $A_{5}$, $A_\nu$ \\
$M_U$ -- $M_{\nu_R}$    & $SU(3)_C \otimes SU(2)_L \otimes U(1)_Y$
                       & $M_1$, $M_2$, $M_3$   \\
             &         & $M^2_{\tilde Q}$, $M^2_{\tilde U}$, $M^2_{\tilde D}$\\
             &         & $M^2_{\tilde L}$,
                         $M^2_{\tilde \nu_R}$, $M^2_{\tilde E}$ \\
             &         & $A_{u}$, $A_{d}$, $A_{\tau}$, $A_\nu$ \\
$M_{\nu_R}$ -- $M_{EW}$ & $SU(3)_C \otimes SU(2)_L \otimes U(1)_Y$
                       & $M_1$, $M_2$, $M_3$   \\
             &         & $M^2_{\tilde Q}$, $M^2_{\tilde U}$, $M^2_{\tilde D}$\\
             &         & $M^2_{\tilde L}$, $M^2_{\tilde E}$ \\
             &         & $A_{u}$, $A_{d}$, $A_{\tau}$ \\
\hline
\end{tabular}
\end{center}
\end{table}
\begin{figure*}
\setlength{\unitlength}{1mm}
\begin{center}
\begin{picture}(160,140)
\put(-4,0){\mbox{\epsfig{figure=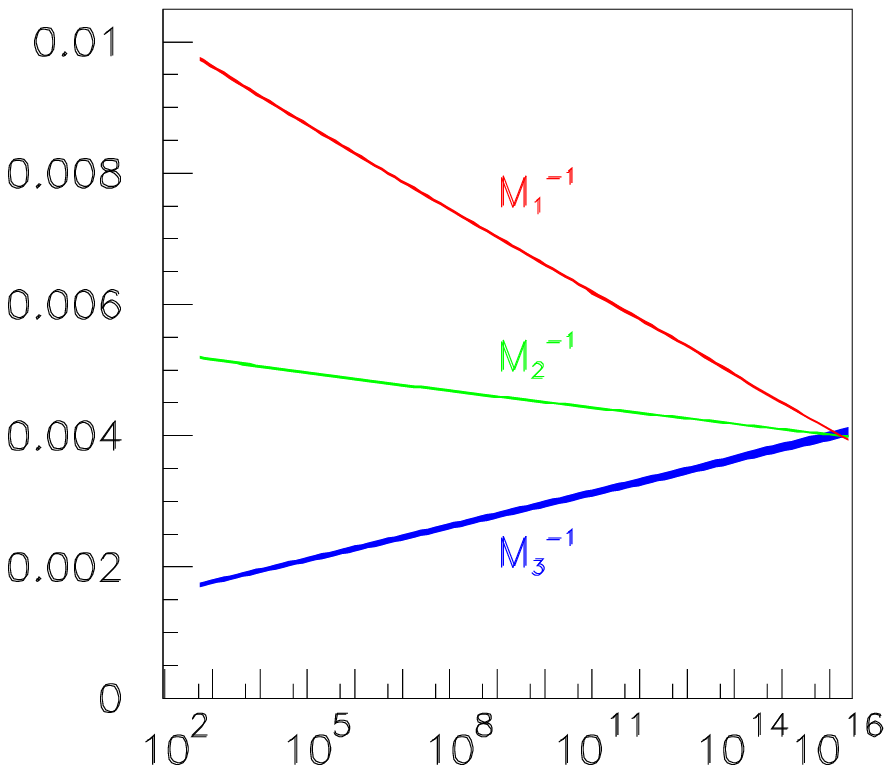,height=17cm,width=18cm}}}
\put(89,87){\mbox{\epsfig{figure=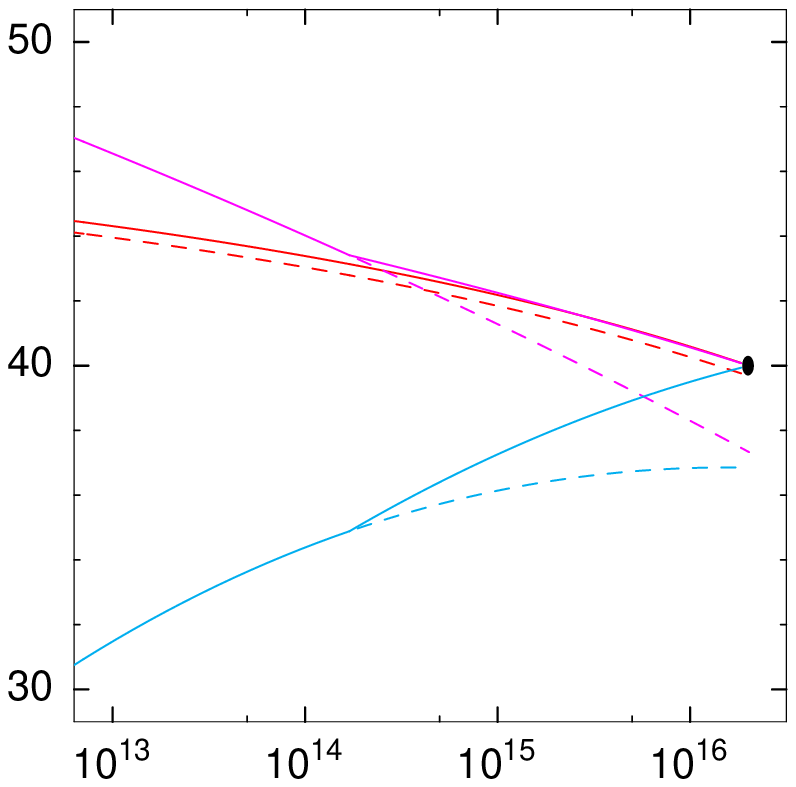,height=6.6cm,width=6.9cm}}}
\put(-4,-86){\mbox{\epsfig{figure=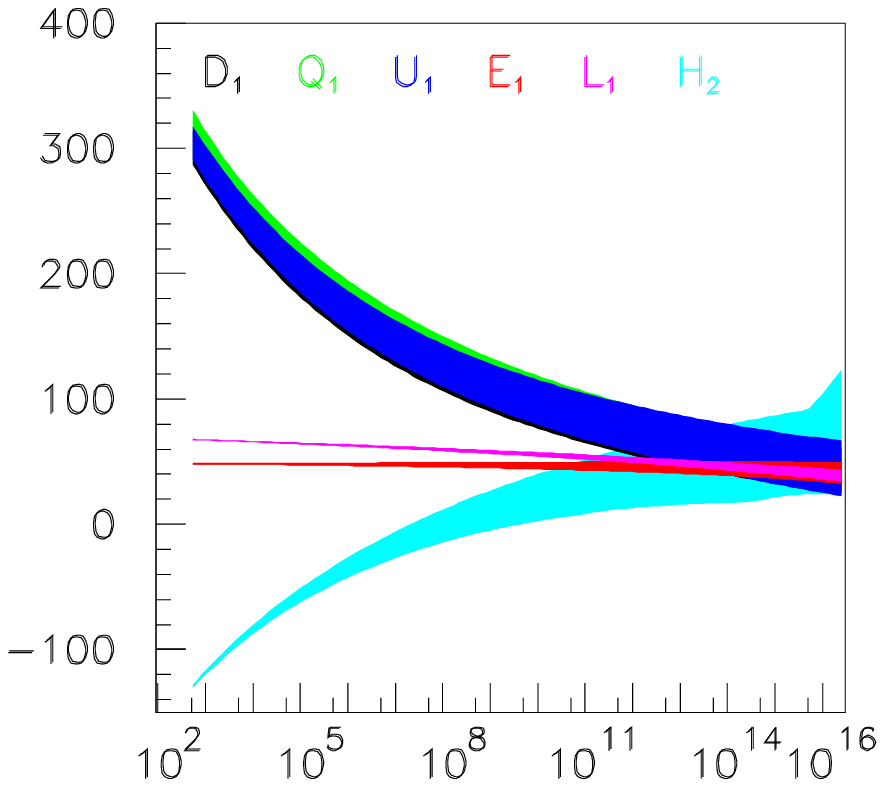,height=17cm,width=18cm}}}
\put(78,-86){\mbox{\epsfig{figure=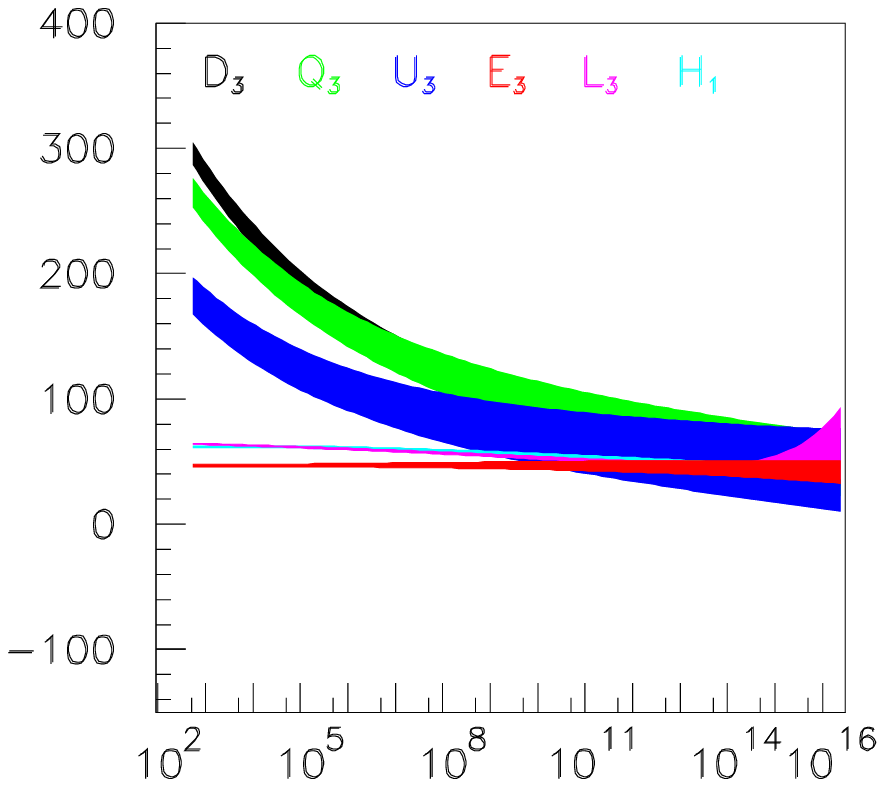,height=17cm,width=18cm}}}
\put(-1,158){\mbox{\bf (a)}}
\put(14,156){\mbox{$1/M_i$~[GeV$^{-1}$]}}
\put(65,82){\mbox{$Q$~[GeV]}}
\put(80,158){\mbox{\bf (b)}}
\put(95,156){\mbox{$M^2_{\tilde j}$~[$10^3$ GeV$^2$]}}
\put(147,83){\mbox{$Q$~[GeV]}}
\put(97,143){\mbox{\small $M^2_{L_3}$}}
\put(98,129){\mbox{\small $M^2_{E_3}$}}
\put(99,106){\mbox{\small $M^2_{H_2}$}}
\put(122.5,100){\mbox{\small $\nu_{R_3}$}}
\put(121.3,100){\vector(0,1){5}}
\put(120,145){\mbox{\small ----- $\nu_R$, $\tilde{\nu}_R$ included}}
\put(120,140){\mbox{\small - - -  $\nu_R$, $\tilde{\nu}_R$ excluded}}
\put(-1,73){\mbox{\bf (c)}}
\put(14,71){\mbox{$M^2_{\tilde j}$~[$10^3$ GeV$^2$]}}
\put(65,-3){\mbox{$Q$~[GeV]}}
\put(80,73){\mbox{\bf (d)}}
\put(95,71){\mbox{$M^2_{\tilde j}$~[$10^3$ GeV$^2$]}}
\put(147,-3){\mbox{$Q$~[GeV]}}
\end{picture}
\end{center}
\caption{{\bf LR--SUGRA with $\nu_R$:} {\it  Evolution of 
(a) gaugino mass parameters,
 (b) evolution of third generation slepton mass parameters and
     Higgs mass parameters  $M^2_{H_2}$;
 (c) evolution of first--generation sfermion mass parameters and
     Higgs mass parameters $M^2_{H_2}$;
 (d) evolution of third generation sfermion mass parameters and
     Higgs mass parameters  $M^2_{H_1}$.
    The mSUGRA point probed is characterized by the
parameters $M_0 = 200$~GeV, 
$M_{1/2} = 250$~GeV, $A_0$ = -100~GeV, $\tan \beta = 10$, 
and $\mathrm{sign}(\mu) = (+)$.
[The widths of the bands indicate the 1$\sigma$ CL.]
}}
\label{fig:sugraRnu}
\end{figure*} 

In this left--right supergravity point, called LR--SUGRA for short\footnote{
Other left--right scenarios will be presented in a
  forthcoming publication.}, we
have probed the same SUSY parameters as above, complemented by the
same universal parameters in the right--handed sneutrino sector.  The
sneutrinos $\tilde \nu_L$ and $\tilde \nu_R$ mix by the (large) Yukawa
interactions in the $\hat{\nu}_R$ sector of the 
superpotential to form the mass eigenstates
$\tilde \nu_1$ and $\tilde \nu_2$. Also in this sector an effective
seesaw mechanism is induced by the large $\nu_R$ mass,
as can be most easily seen by considering the one generation case:
\begin{eqnarray}
\label{eq:SneuMatrix}
&&m^2 = 
\left( 
\begin{array}{cc}
  M^2_{\tilde L} + \frac{1}{2} m^2_Z \cos 2 \beta 
& \frac{1}{\sqrt{2}} Y_\nu (A_\nu v_2 - \mu v_1 ) \\
\frac{1}{\sqrt{2}} Y_\nu (A_\nu v_2 - \mu v_1 )
&   M^2_{{\tilde\nu}_R} + M^2_{\nu_R}
\end{array}
\right)
\end{eqnarray}
In this mass matrix $M_{\nu_R}$ is the [GUT-scale] mass of the 
right--handed neutrino, 
$M_{{\tilde\nu}_R}$ the scalar [TeV-scale] mass parameter of the right sneutrino, 
$Y_\nu$ and
$A_\nu$ the neutrino Yukawa coupling and the neutrino trilinear coupling,
respectively. $v_1$ and $v_2$ are the vacuum expectation values of the
Higgs field with isospin $-1/2$ and isospin $1/2$, respectively.
The approximate eigenvalues of the sneutrino mass matrix read 
\begin{eqnarray}
  \label{eq:SnuMass}
  m^2_{{\tilde \nu}_1} &\simeq& M^2_{\tilde L} + \frac{1}{2} m^2_Z \cos 2 \beta
       -  Y^2_\nu \frac{(A_\nu v_2 - \mu v_1 )^2}{2 M^2_{\nu_R}} \\
 m^2_{{\tilde \nu}_2} &\simeq& M^2_{\nu_R} + M^2_{{\tilde\nu}_R}
       +  Y^2_\nu \frac{(A_\nu v_2 - \mu v_1 )^2}{2 M^2_{\nu_R}} 
\end{eqnarray}
Therefore,
to a very good approximation, $\tilde \nu_1$ coincides with $\tilde \nu_L$ 
and $\tilde \nu_2$ with $\tilde \nu_R$.
The heavy right--handed neutrino
masses are calculated by identifying the Yukawa couplings with the
up-type quark couplings in the quark sector at the GUT scale
(largely equivalent to the $SO(10)$ scale in this regard) and by
identifying the light neutrino masses with the neutrino mass differences
in the large mixing angle for the solar neutrino problem solution:  $m_{\nu_{L1}}= 10^{-5}$~eV, 
$m_{\nu_{L2}}= 3 \cdot 10^{-3}$~eV, $m_{\nu_{L3}}= 6 \cdot 10^{-2}$~eV;
$M_{\nu_{R_1}} = 3 \cdot 10^9$~GeV,  
$M_{\nu_{R_2}} = 1.4 \cdot  10^{11}$~GeV, 
$M_{\nu_{R_3}} = 1.7 \cdot 10^{14}$~GeV.

The impact on the evolution of the mass parameters 
is rather simple. In the analysis of the first two
generations the Yukawa interactions involving the heavy neutrinos and
the R-sneutrinos are so small that their effect is not noticeable in
practice. The evolution of the gaugino and scalar mass parameters is
not affected by the left--right extension of the system in the present
form  as is evident from Figs.~\ref{fig:sugraRnu}(a)
and (c).  This is only different for the third generation and for
$M^2_{H_2}$ owing to the
enhanced Yukawa coupling in this case as shown in Figs.~\ref{fig:sugraRnu}(b) 
and (d). The sensitivity
to the intermediate $\nu_R$ scales is rather weak because  
neutrino Yukawa couplings affect the evolution of the sfermion mass parameters
only mildly.
Since the $\nu_R$ of the third generation is unfrozen only beyond
$Q=M_{\nu_R}$ the impact of the LR extension becomes visible in the evolution
only at very high scales.
In \fig{fig:sugraRnu}(b) we display the evolution of $M^2_{\tilde E_3}$,
$M^2_{\tilde L_3}$ and $M^2_{H_2}$
for illustrative purposes.  The full lines
include the effects of the right--handed neutrino, which are
 to be compared with the dashed
lines where the $\nu_R$ effects are removed. 
The scalar mass parameter $M^2_{\tilde E_3}$ appears unaffected by the 
right--handed sector, while $M^2_{\tilde L_3}$ and $M^2_{H_2}$ clearly are.
Only the picture including $\nu_R$, $\tilde{\nu_R}$ is compatible
with the unification assumption.
The kinks in the
evolution of $M^2_{{\tilde L}_3}$ and $M^2_{\tilde {H_2}}$ can be traced back
 to the
fact that around $10^{14}$ GeV the third generation (s)neutrinos 
become quantum mechanically effective, given a large enough neutrino
Yukawa coupling
to influence the evolution of these mass parameters.
A much better understanding of the third generation family must be
achieved to draw quantitative conclusions beyond the rough estimates 
of the  $\nu_R$ scales sketched in the present analysis.

\section{Gauge Mediated Supersymmetry Breaking}

\begin{figure*}
\setlength{\unitlength}{1mm}
\begin{center}
\begin{picture}(160,140)
\put(-4,0){\mbox{\epsfig{figure=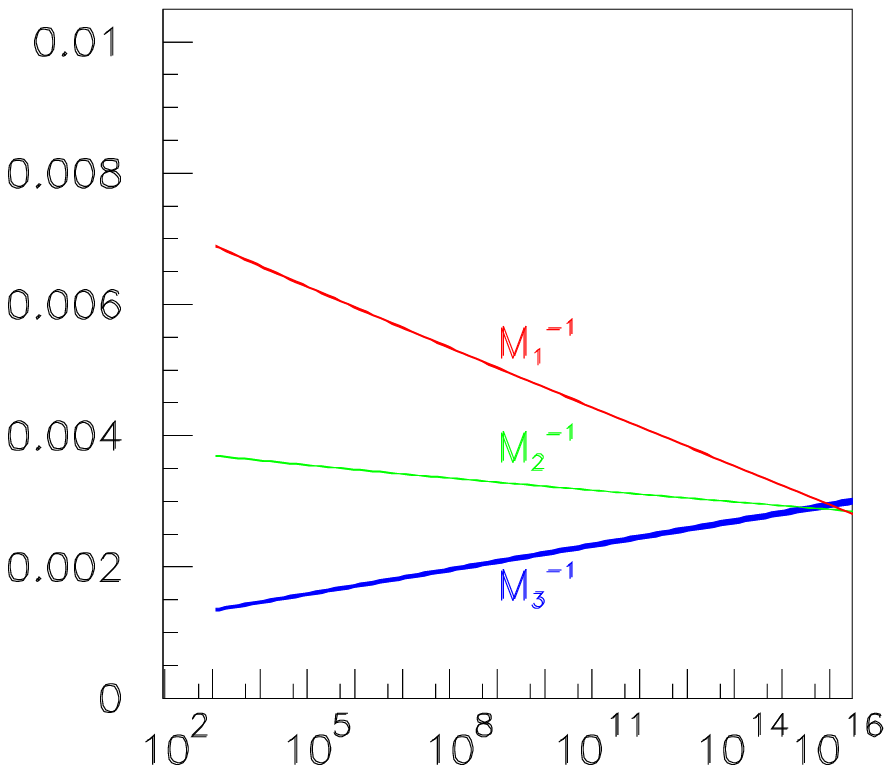,height=17cm,width=18cm}}}
\put(80,81){\mbox{\epsfig{figure=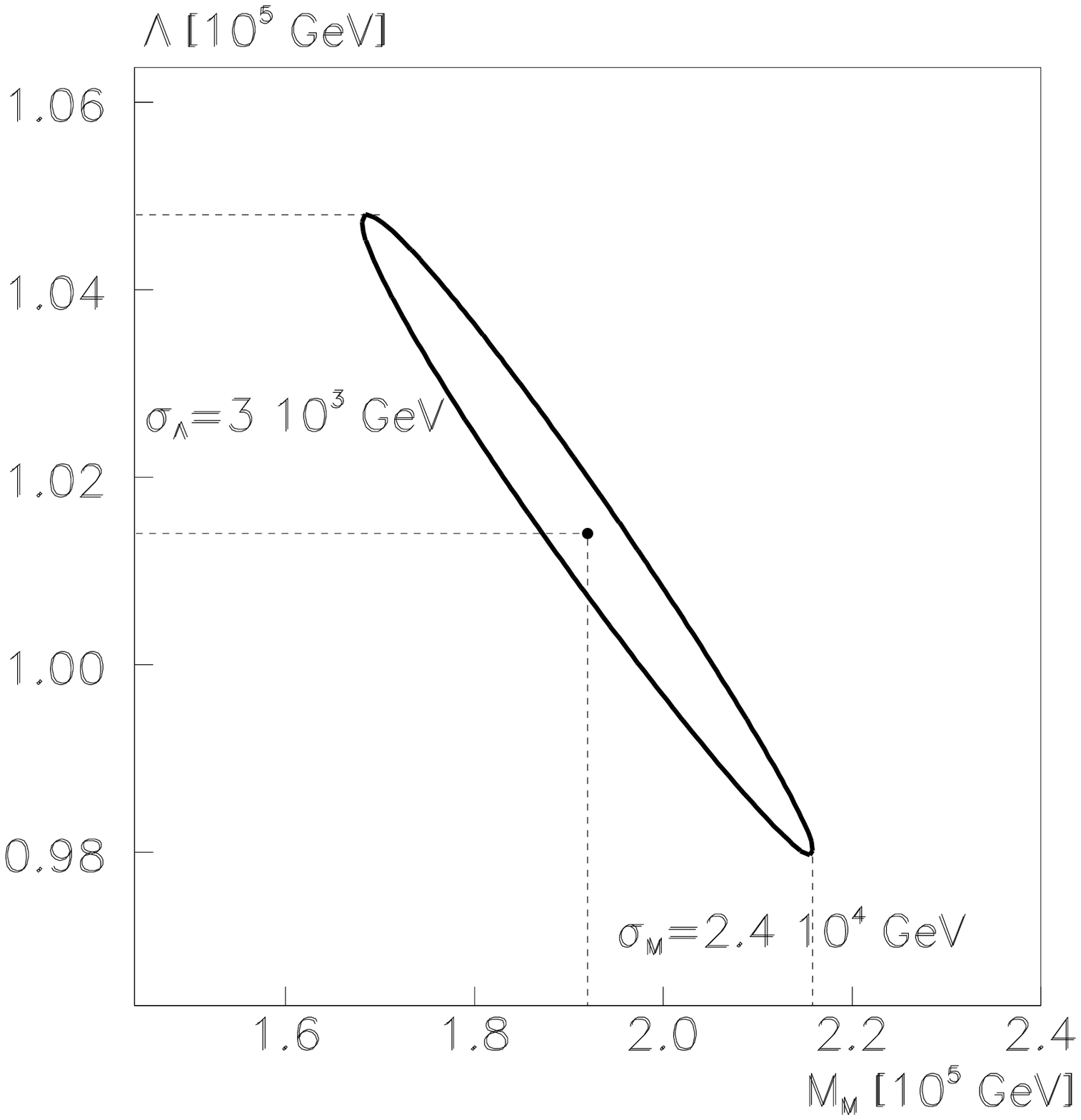,
                                   height=8cm,width=8.7cm}}}
\put(-4,-86){\mbox{\epsfig{figure=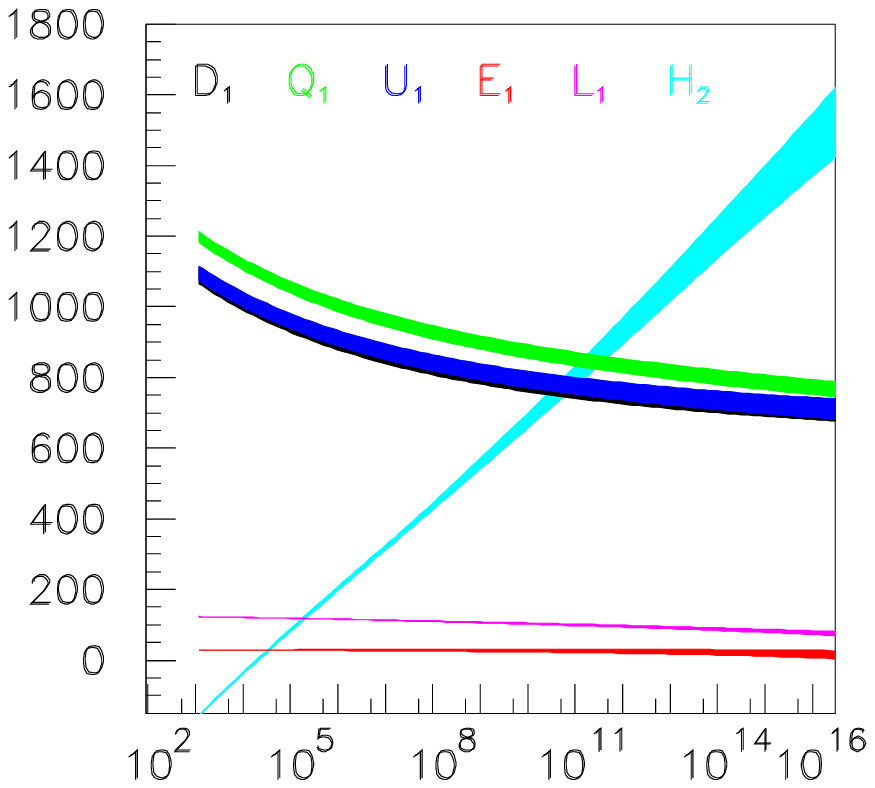,height=17cm,width=18cm}}}
\put(78,-86){\mbox{\epsfig{figure=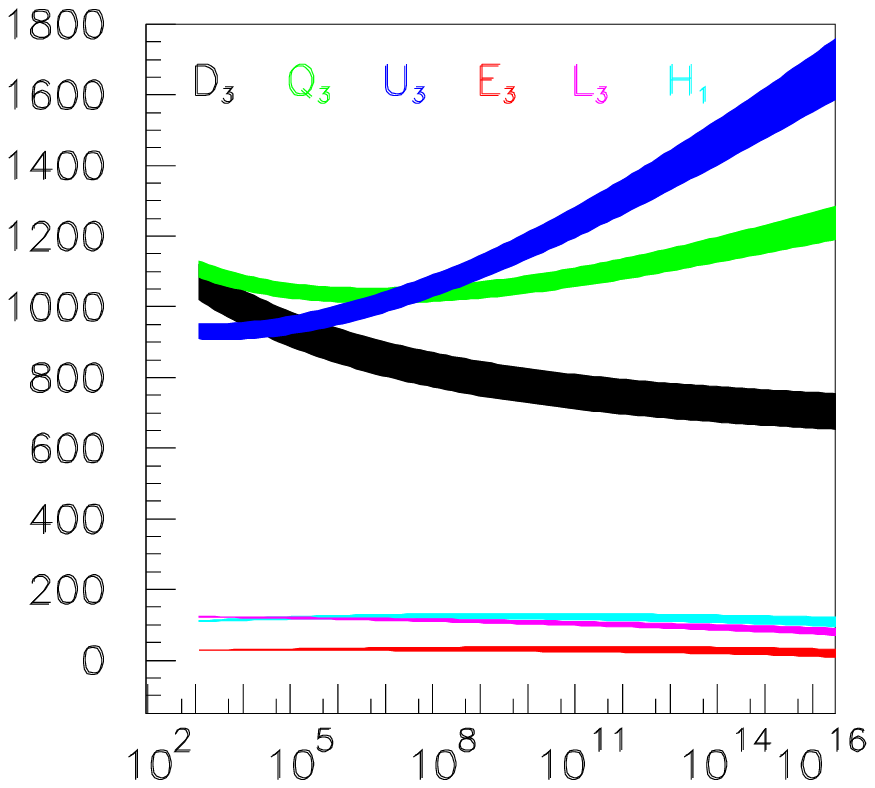,height=17cm,width=18cm}}}
\put(-1,158){\mbox{\bf (a)}}
\put(14,156){\mbox{$1/M_i$~[GeV$^{-1}$]}}
\put(65,82){\mbox{$Q$~[GeV]}}
\put(80,159){\mbox{\bf (b)}}
\put(-1,73){\mbox{\bf (c)}}
\put(14,71){\mbox{$M^2_{\tilde j}$~[$10^3$ GeV$^2$]}}
\put(65,-3){\mbox{$Q$~[GeV]}}
\put(27,24.5){$M_M$}
\put(28.5,24){\vector(0,-1){5}}
\put(80,73){\mbox{\bf (d)}}
\put(95,71){\mbox{$M^2_{\tilde j}$~[$10^3$ GeV$^2$]}}
\put(147,-3){\mbox{$Q$~[GeV]}}
\end{picture}
\end{center}
\caption{{\bf GMSB:} {\it  Evolution of 
(a) gaugino mass parameters;
 (b) $\Lambda$--$M_M$ determination
 in the
bottom--up approach.
 (c) first--generation sfermion mass parameters and 
     Higgs mass parameter $M^2_{H_2}$;
 (d) third--generation sfermion mass parameters and 
     Higgs mass parameter  $M^2_{H_1}$;
The point probed, SPS\#8, is characterized by the
parameters $M_M = 200$~TeV, 
$\Lambda = 100$~TeV, $N_5 = 1$, $\tan \beta = 15$, 
and $\mathrm{sign}(\mu) = (+)$.
[The widths of the bands indicate the 1$\sigma$ CL.]
}}
\label{fig:Gmsb}
\end{figure*} 

Motivated by the observed suppression of flavour changing neutral
transitions, supersymmetry breaking mediated by gauge interactions 
from a secluded sector to the visible eigen--world,
offers an automatic solution to this problem \cite{gmsb,gmsb2}.
The scalar and the F components of a Standard--Model singlet superfield $S$
acquire vacuum expectation values $\langle S \rangle$ and 
$\langle F_S \rangle $ through interactions with
fields in the secluded sector, thus breaking supersymmetry.\footnote{
A solution of the doublet--triplet splitting problem can be found in GMSB
by introducing two different $S$ fields. The masses of supersymmetric particles
are less constrained in this approach than in the one-scale model, and they
depend on the values of the two $\Lambda_i$ parameters. In particular,
the approximate equality of the gaugino masses at the GUT scale is
lifted, see \cite{Witten02} for details.}
Vector-like messenger fields $M$, carrying non--zero 
$SU(3)\times SU(2) \times U(1)$ charges and coupling to $S$, transport
the supersymmetry breaking to the eigen--world.
The system is characterized by the mass $M_M \sim$ $\langle S \rangle $ 
of the messenger
fields and the mass 
scale $\Lambda = {\langle F_S \rangle} / {\langle S \rangle} $ setting the seize 
of the gaugino and scalar masses.  $M_M$ is expected to be in the
range of 10 to $10^6$ TeV and $\Lambda$ has to be smaller than $M_M$. 
The gaugino masses  
\begin{equation}
M_i(M_M) = (N_5+3 N_{10}) g\left(\Lambda/M_M\right)
\alpha_i(M_M) \Lambda
\label{bc1}
\end{equation}
are generated by loops of scalar and fermionic messenger component
fields; $N_i$ is the multiplicity of  messengers in the
$5+\overline{5}$ and $10+\overline{10}$ vector-like multiplets, and
\begin{eqnarray}
g(x) &=&{1+x\over x^2}\log(1+x)  +   (x\rightarrow-x)
\end{eqnarray}
is the messenger--scale threshold function \cite{Martin} which
approaches unity for $\Lambda \ll M_M$. Masses of the scalar fields in the
visible sector
are generated by 2-loop effects of gauge/gaugino and messenger fields:
\begin{equation}
M^2_{\tilde j} (M_M) = 2  (N_5 + 3 N_{10}) f\left(\Lambda/M_M\right)
\sum_{i=1}^3  k_i   C_i^j \alpha^2_i(M_M) \Lambda^2
\label{bc2}
\end{equation}
with $k_i=1,1,3/5$ for $SU(3)$, $SU(2)$, and $U(1)$, respectively;
the coefficients $C_i^j$ are the quadratic Casimir invariants,
being 4/3, 3/4, and $Y^2/4$ for the fundamental representations ${\tilde j}$ 
in the groups $i = SU(3), SU(2)$ and $U(1)$, with $Y=2(Q-I_3)$ denoting 
the usual hypercharge; also
the threshold function \cite{Martin} 
\begin{eqnarray}
f(x)&=& {1+x\over x^2}\biggl[\log(1+x) -2{\rm Li}_2
\left({x\over1+x}\right)+\ {1\over2}{\rm Li}_2
\left({2x\over1+x}\right)\biggr] 
  +   (x\rightarrow-x)
\end{eqnarray} 
approaches unity for $\Lambda \ll M_M$.
As evident from \eq{bc2} scalar particles with identical Standard--Model 
charges squared have equal
masses at the messenger scale $M_M$.
In the minimal version of GMSB, the $A$ parameters
are generated at 3-loop level and they are practically zero at $M_M$.

We have investigated this scheme for the point $\Lambda = 100$~TeV,
$M_M = 200$~TeV, $N_5=1$, $N_{10}=0$, $\tan\beta=15$ and $\mu>0$
corresponding to the Snowmass Point SPS\#8. 
We find for the low energy data: BR($b\to s \gamma) = 3.7 \cdot 10^{-4}$, 
$\Delta[g - 2]_\mu = 15 \cdot 10^{-10}$, $\Delta \rho = 64 \cdot 10^{-5}$.
The evolution\footnote{The same formulae as in 
\app{sec:AppRGEGUT} apply for the GMSB boundary conditions 
at the electroweak scale 
by replacing $M_U$ by $M_M$, the GMSB scale.}
 of the gaugino and sfermion mass parameters
of the first and third generation as well as the Higgs mass parameters, 
including 2-loop $\beta$--functions,
are presented in \fig{fig:Gmsb}. 
Owing to the influence of the $A$--parameters in the 2-loop
RGEs for the gaugino mass parameters,  the gaugino mass
parameters do not meet at the same point as the gauge couplings in this
scheme.
It is obvious from the figure that the GMSB scenario cannot be confused
with the universal supergravity scenario\footnote{A comparison of the mass 
characteristics at the low scale between mSUGRA and GMSB in a top-down
approach is presented in Ref.\cite{Bagg}.}. [Specific experimental
signatures generated in the decays of the next to lightest supersymmetric
particle, the neutralino $\tilde \chi^0_1$ or the stau $\tilde \tau_1$,
to gravitinos which are very light in GMSB, provide a complementary 
experimental discriminant, see \cite{Aguilar-Saavedra:2001rg,blair}].

The bands of the slepton $L$--doublet mass parameter $M^2_{\tilde L}$ and the
Higgs parameter $M^2_{H_2}$, which carry the same moduli of
standard--model charges, cross at the scale $M_M$. The crossing,
which is indicated by an arrow in \fig{fig:Gmsb}(c), is a necessary
condition (in the minimal form) for the GMSB scenario to be realized. 

The two scales $\Lambda$ and $M_M$ can be extracted from the spectrum
of the gaugino and scalar particles.  Combining the two species allows
to determine the multiplicity coefficient $(N_5 + 3 N_{10})$ in
addition.  The dependence of the spectra on $\Lambda$ is, trivially,
very strong. The messenger scale $M_M$ can be determined only from the
running of the masses between the messenger scale and the electroweak
scale ; despite being governed by weakly varying logarithms, the
accuracy in determining $M_M$ is surprisingly good.  For the point
analyzed in the example above, the following accuracy for the mass
parameters and the messenger
multiplicity has been found:\\
\begin{eqnarray}
\Lambda &=& (1.01 \pm 0.03) \cdot 10^2 \; \rm {TeV}\\
M_M &=&(1.92 \pm 0.24) \cdot 10^2 \; \rm {TeV} \\
N_5 + 3 N_{10} &=& 0.978 \pm 0.056
\end{eqnarray}
The correlation between $\Lambda$ and $M_M$ is shown in \fig{fig:Gmsb}(b).

\begin{table}[t]
\caption[]{\it Average ratios of the scalar mass parameters as predicted
               in GMSB solely by group factors and gauge couplings.} 
\label{tab:gmsb_ratios}
\begin{center}
\begin{tabular}{c||c|c}
Mass$^2$ Ratios & $<$GMSB$>$& $\ne$mSUGRA  \\ \hline \hline
$H_2^2/L_1^2 $ & 1 &  ---  \\ \hline
$E_1^2/L_1^2 $ & 0.25 &  $\ge$0.8  \\
$Q_1^2/L_1^2 $ & 8.87 &  $\le $3.2  \\
$U_1^2/L_1^2 $ & 8.03 & $\le$ 3.0  \\
$D_1^2/L_1^2 $ & 7.95 &  $\le$3.2  \\
$H_1^2/L_1^2 $ & 1 & $\le$1.0    \\ 
\hline
\end{tabular}
\end{center}
\end{table}
\begin{figure*}
\setlength{\unitlength}{1mm}
\begin{center}
\begin{picture}(160,78)
\put(-2,-7){\mbox{\epsfig{figure=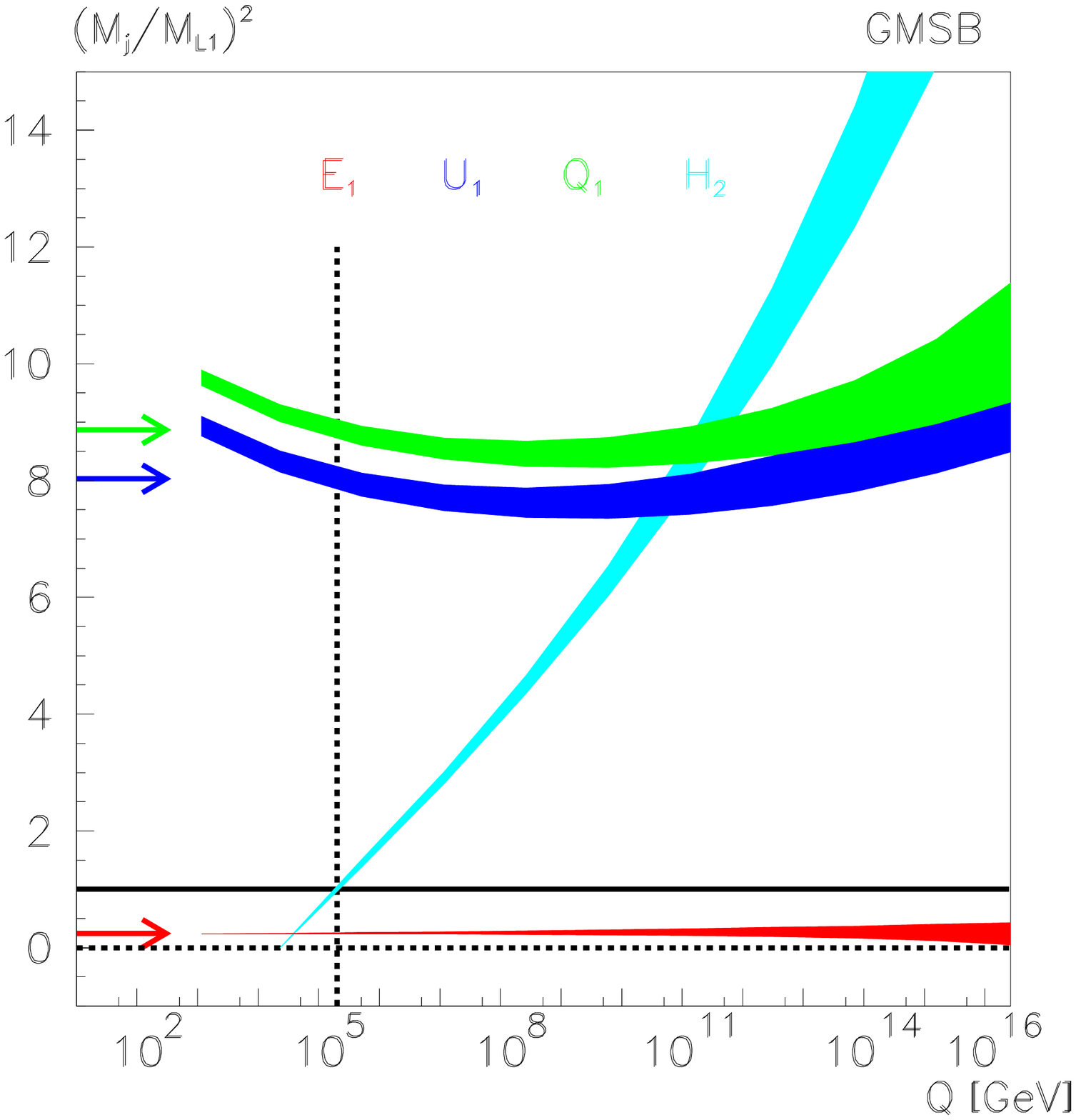,
                                   height=8.5cm,width=8.5cm}}}
\put(82,-7){\mbox{\epsfig{figure=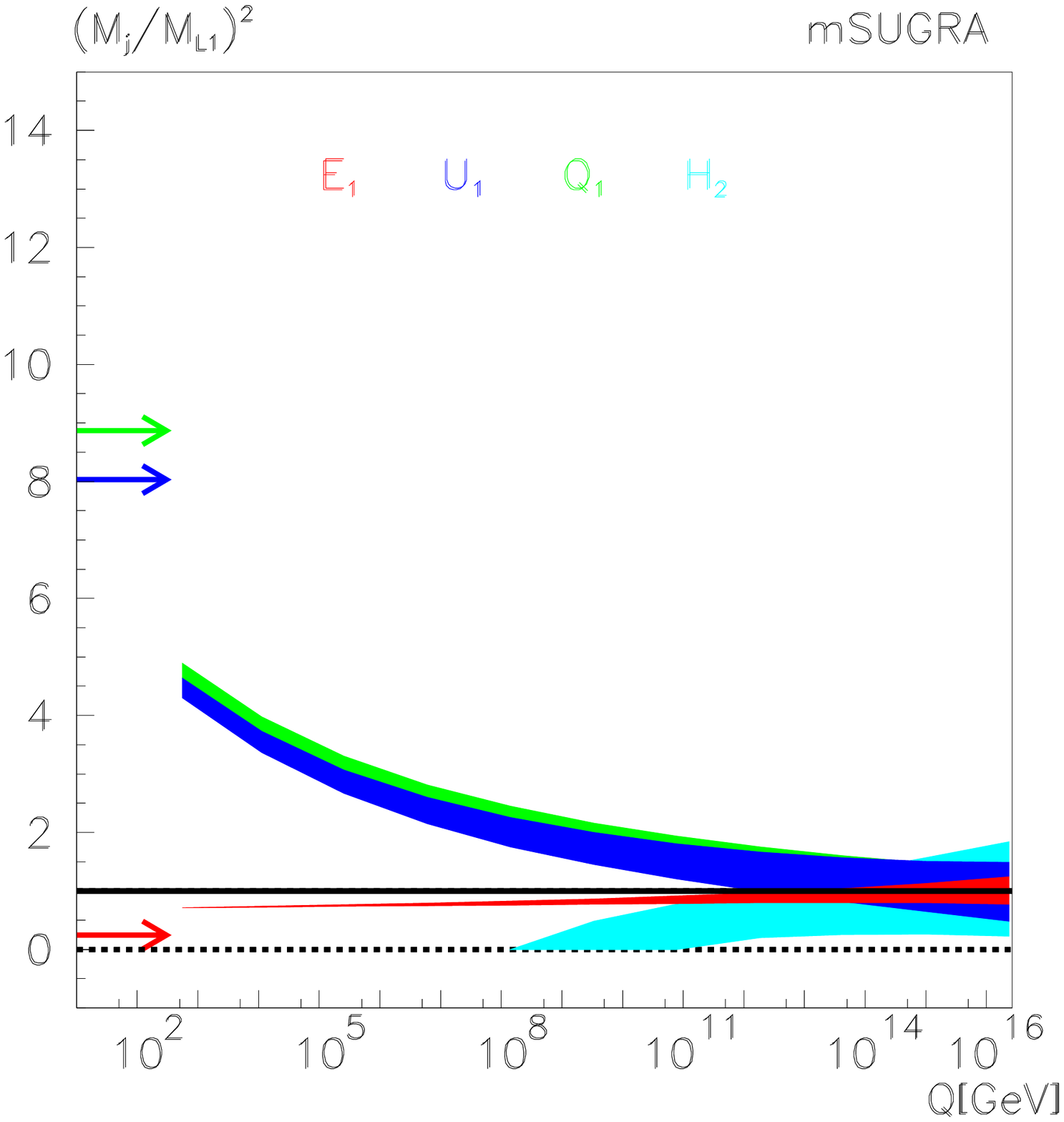,
                                   height=8.5cm,width=8.5cm}}}
\put(-2,73){\mbox{\bf a)}}
\put(82,73){\mbox{\bf b)}}
\end{picture}
\end{center}
\caption{\it Evolution of representative ratios of first generation scalar 
          masses squared, a) in case of GMSB; b) in case of mSUGRA. The
          messenger scale $M_M$ is defined as the scale where $M_{H_2} =
          M_{L_1}$. The arrows in \underline {both} figures indicate 
          the expectation values of the mass ratios squared in GMSB 
          at the scale $M_M$.}
\label{fig:ratio}
\end{figure*} 

While the gaugino masses in GMSB evolve nearly in the same way as in mSUGRA,
the running of the scalar masses is quite different in both theories. 
Moreover, at the messenger scale the ratios of scalar masses squared in
the simplest version of 
GMSB are determined solely by group factors and gauge couplings, being 
independent of the specific GMSB characteristics, i.~e. messenger 
multiplicities and $\Lambda$ mass scale:\\
\begin{equation}
M^2_{\tilde j} (M_M)/M^2_{\tilde j^\prime} (M_M) = 
\sum_{i=1}^3  k_i   C_i^j \alpha^2_i(M_M)/
\sum_{i=1}^3  k_i   C_i^{j^{\prime}} \alpha^2_i(M_M).
\label{bc3}
\end{equation}
The predictions for these ratios are listed in \tab{tab:gmsb_ratios}. 
The ratios
in GMSB are distinctly different from the ratios in mSUGRA, taken at the scale
where the upper boundary of the $2\sigma$ band for $H^2_2/L^2_1$ approaches 
unity from below. [Ideally all ratios approach unity only at the grand unification scale $M_U$ in mSUGRA.] The distinct differences between GMSB and mSUGRA
are clearly visible in Figs.~\ref{fig:ratio}(a) {\it vs.} (b).   

\section{String Induced Supersymmetry Breaking}

In the previously analyzed SUGRA and GMSB models the structure
of the supersymmetry breaking mechanisms {\it sui generis} 
and the fields involved in the
hidden sectors are shielded from the eigen--world. Four--dimensional
strings naturally give rise to a minimal set of fields for inducing
supersymmetry breaking; they play the r\^ole of the fields in the hidden
sectors: the dilaton $S$ and the moduli $T_m$ chiral
superfields which are generically present in large classes of
4--dimensional heterotic string theories\footnote{For other scenarios
see Ref.\cite{All}.}. The vacuum expectation
values of $S$ and $T_m$, generated by genuinely non--perturbative
effects, determine the soft supersymmetry breaking parameters.
In this approach, grand unification at the standard scale can be
reconciled with the higher string scale by moduli dependent string loop
corrections. 

In the following we assume that all moduli fields get the same vacuum
expectation values and that they couple in the same way to matter fields.
Therefore, we omit the index $m$ and take only one
moduli field $T$.
 The properties of the supersymmetric theories
are quite different for dilaton and moduli dominated scenarios. This
can be quantified by introducing a mixing angle $\theta$,
characterizing the $\tilde S$ and $\tilde T$ wave functions of the
Goldstino, which is associated with the breaking of supersymmetry and
which is absorbed to generate the mass of the gravitino:
\begin{eqnarray}
\tilde G &=& \sin \theta \, \tilde{S} + \cos \theta \, \tilde{T}
\end{eqnarray}
  The mass
scale is set by the second parameter of the theory, the gravitino mass
$m_{3/2}$.

A dilaton dominated scenario, i.e.~$\sin \theta \to 1$, leads to
universal boundary conditions of the soft supersymmetry breaking
parameters.  Universality is broken\footnote{For other mechanisms 
of breaking universality see e.g. Ref.\cite{Gun}.}  only slightly 
by small loop
effects. On the other hand, in moduli field dominated scenarios,
$\cos\theta \to 1$, the gaugino mass parameters are universal to lowest
order [broken only in higher orders], but universality is not realized
for the scalar mass parameters in general. The breaking is quantified 
by modular weights $n_j$ characterizing the couplings between the
matter and the moduli fields in orbifold compactifications.
Within one generation significant
differences between left and right field components and between
sleptons and squarks can occur; since these patterns are
modified only by small loop effects between different generations,
flavour--changing neutral effects remain suppressed.

In leading order, and next--to--leading order parameterized by
the quantities $\Delta M$, the masses \cite{Binetruy:2001md}
are given by the following expressions for
the gaugino sector, 
\begin{eqnarray}
M_i &=&  - g_i^2 m_{3/2} s {\sqrt{3} \sin \theta} + \Delta M_i \\
 \Delta M_i &=& -g_{i}^{2}  m_{3/2} \left\{
b_{i} + 
   {s \sqrt{3}\sin\theta} 
 g_{s}^{2}\left(C_i
-\sum_{j}C_{i}^{j}\right) \right.
 \nonumber \\ &&  + \left. 2 \, t
\cos\theta\,  G_{2}(t)
    \left[ \delta_{\rm GS} + b_{i}
           - 2  \sum_{j}C_{i}^{j} (1+n_j)   \right]
 \right\} / {16\pi^2}
\end{eqnarray}
and for the scalar sector,
\begin{eqnarray}
M_{\tilde j}^2 &=& m^2_{3/2} \left(
  1 + n_j \cos^2 \theta \right) + \Delta M^2_{\tilde j} \\
\Delta M^2_{\tilde j}&=&m^2_{3/2}
 \Bigg\{ \gamma_{j} + 2 t \cos \theta \, G_2(t)
  \sum_{km} \gamma^{km}_j (n_j + n_k + n_m+3) 
 \nonumber \\ && \hspace*{0.8cm} + 
   {2 \sqrt{3} s \sin\theta} 
 \left[ \sum_{i} \gamma_{j}^{i} g_{i}^{2} -
 \frac{1}{2s}
   \sum_{km} \gamma_{j}^{km} \right]
 \Bigg\}
\end{eqnarray}
while the $A$ parameters read 
\begin{eqnarray}
A_{jkm}&=& - m_{3/2} \bigg[ 2 t
 \cos\theta  (n_j + n_k + n_m + 3) G_2(t) 
 - \frac{ \sin \theta }{\sqrt{3}} 
 \bigg]  + \Delta A_{jkm}\\
\Delta A_{jkm}  &=& m_{3/2} (\gamma_j + \gamma_k + \gamma_m)
\end{eqnarray}
The mass $m_{3/2}$ is the gravitino mass introduced earlier.
The gravitino mass can be expressed in terms of the K\"ahler potential
$K$ and the superpotential $W$, which include the (non-perturbative)
solutions of all the fields at the string scale:
$m_{3/2}=\left<\exp(K/2) \overline{W} \right>$.  
$s = \langle S \rangle$ is the vacuum expectation values of the dilaton 
field.
$t = \langle T \rangle / m_{3/2}$ is the vacuum
expectation value of the moduli field(s), and 
$G_2(t) = 2\zeta(t) + 1/2t$ is the 
non-holomorphic Eisenstein function with
$\zeta$ denoting the Riemann zeta function.
$\delta_{GS}$ is the parameter 
of the Green-Schwarz counterterm.
$\gamma_j$ are the anomalous 
dimensions of the matter
fields, the $\gamma^i_j$ and $\gamma^{km}_j$ are their gauge and Yukawa parts,
respectively.
$C_i$, $C^j_i$ are the quadratic Casimir operators for the gauge group 
$G_i$, respectively, in the adjoint representation and in the matter 
representation. 

\begin{figure*}
\setlength{\unitlength}{1mm}
\begin{center}
\begin{picture}(160,140)
\put(2,82){\mbox{\epsfig{figure=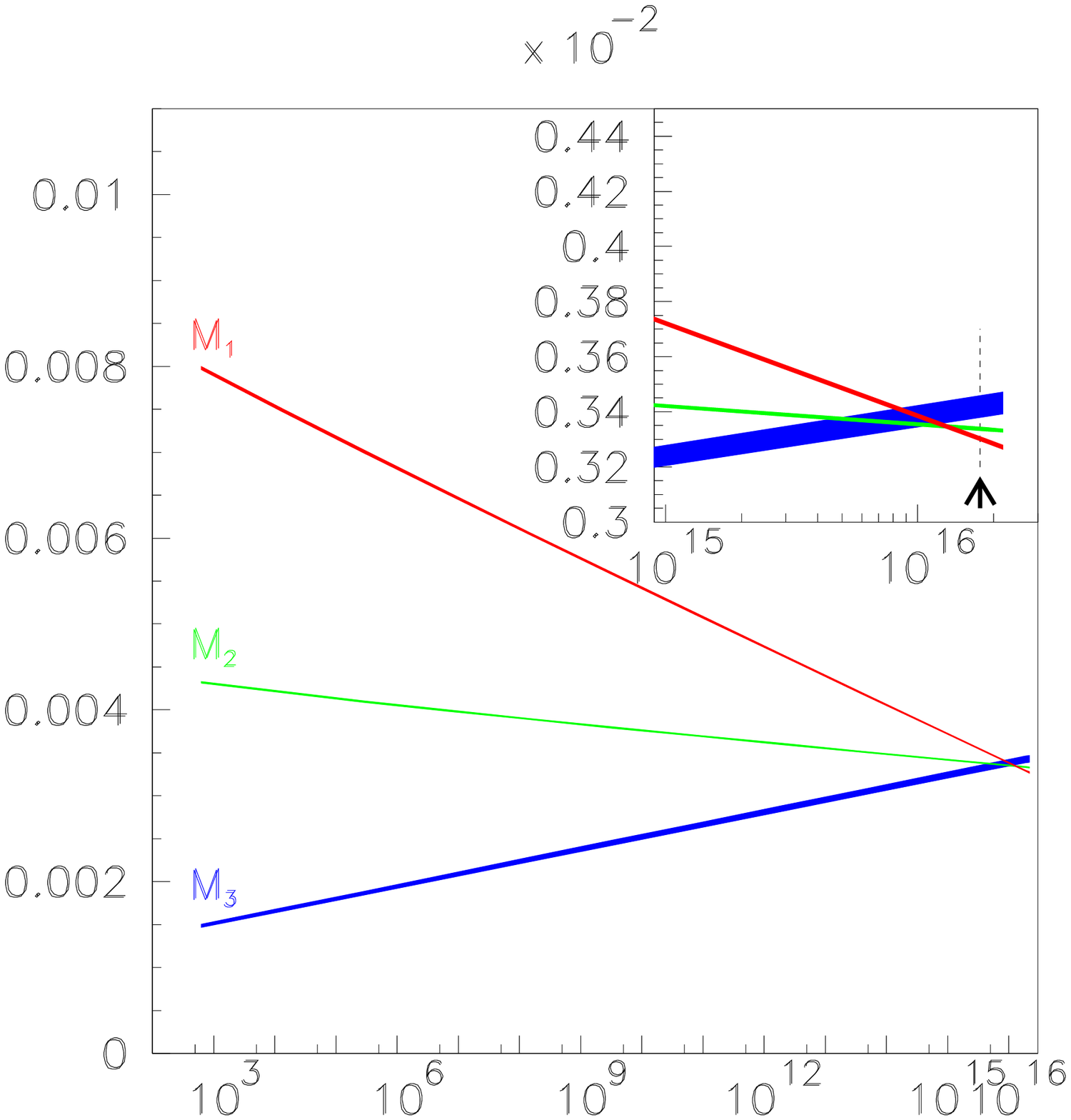,
                                   height=7.9cm,width=8.5cm}}}
\put(80,77){\mbox{\epsfig{figure=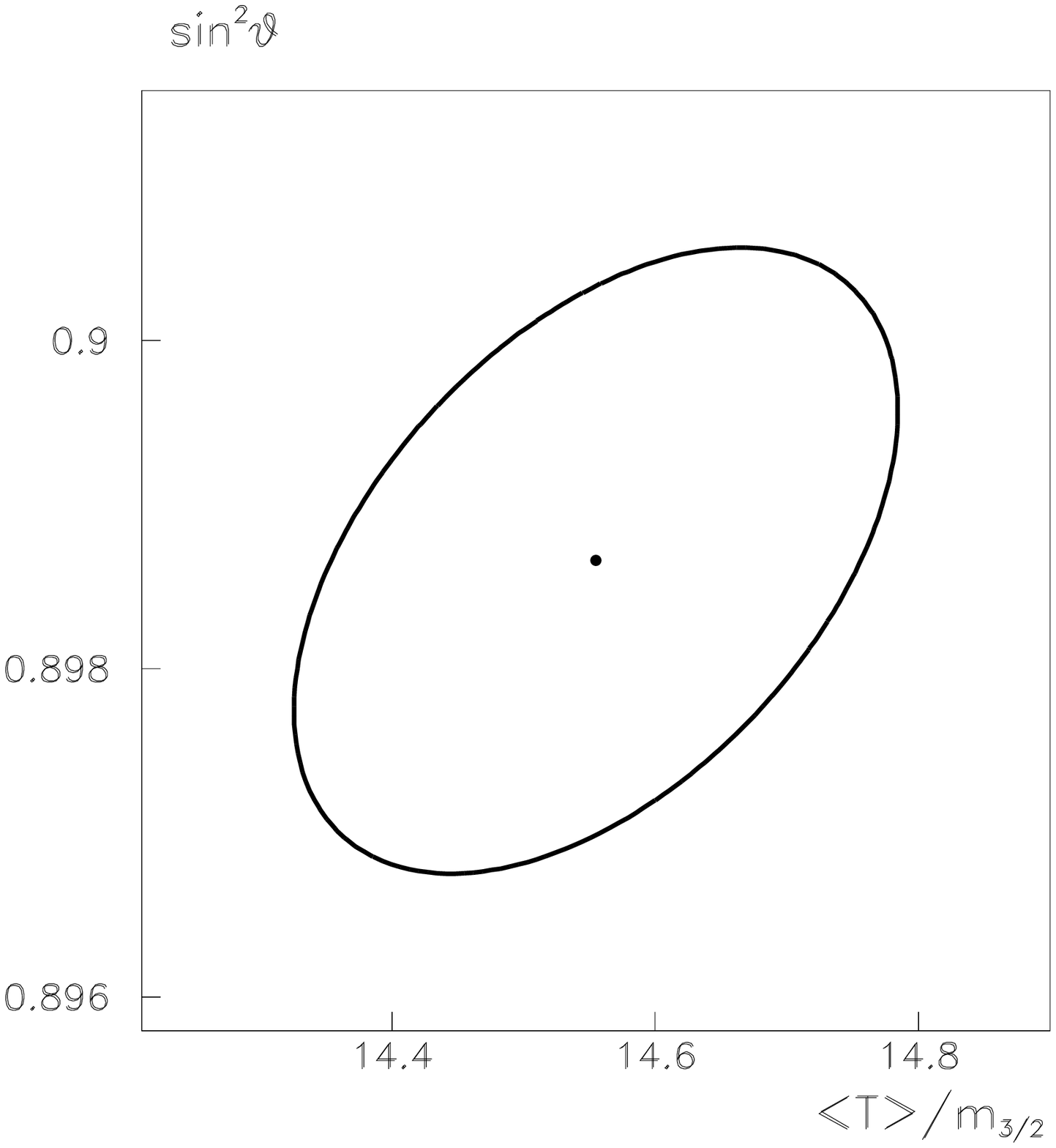,
                              height=8.4cm,width=8.7cm}}}
\put(-4,-86){\mbox{\epsfig{figure=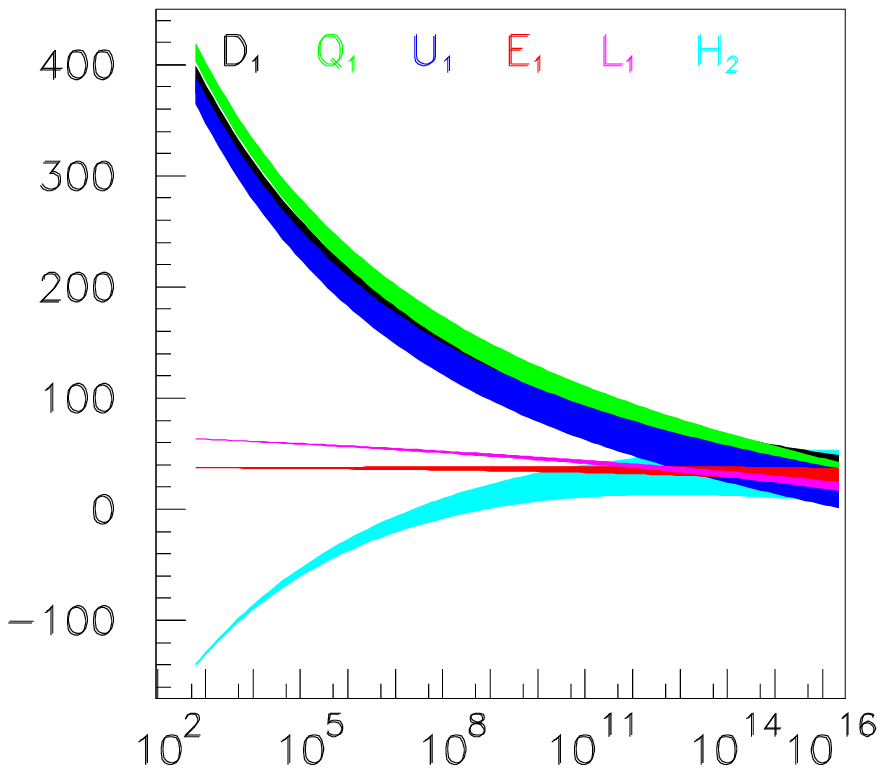,height=17cm,width=18cm}}}
\put(78,-86){\mbox{\epsfig{figure=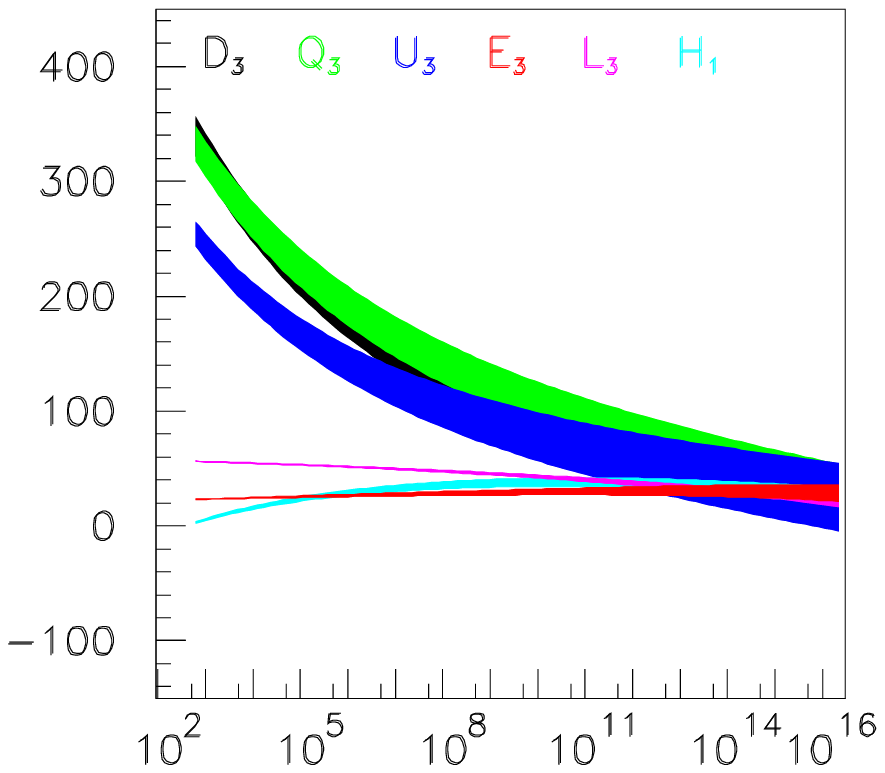,height=17cm,width=18cm}}}
\put(-1,158){\mbox{\bf (a)}}
\put(9,156){\mbox{$-1/M_i$~[GeV$^{-1}$]}}
\put(65,82){\mbox{$Q$~[GeV]}}
\put(80,159){\mbox{\bf (b)}}
\put(-1,73){\mbox{\bf (c)}}
\put(14,71){\mbox{$M^2_{\tilde j}$~[$10^3$ GeV$^2$]}}
\put(65,-3){\mbox{$Q$~[GeV]}}
\put(80,73){\mbox{\bf (d)}}
\put(95,71){\mbox{$M^2_{\tilde j}$~[$10^3$ GeV$^2$]}}
\put(147,-3){\mbox{$Q$~[GeV]}}
\end{picture}
\end{center}
\caption{{\bf String Scenario:} {\it  Evolution of 
 (a) gaugino mass parameters [the insert expands on the breaking
     of universality at the GUT scale];
 (b) correlation between the mixing parameter $\sin^2{\theta}$ 
     and the vacuum expectation value of the moduli field $\langle T
     \rangle$;
 (c) evolution of first--generation sfermion mass parameters and
     Higgs mass parameters $M^2_{H_2}$;
 (d) evolution of third generation sfermion mass parameters and
     Higgs mass parameters  $M^2_{H_1}$.
The point probed is characterized by the
parameters  $m_{3/2}=180$~GeV, $\delta_{GS}=0$, $\sin^2 \theta=0.9$,
 $\langle T \rangle = 14\,m_{3/2}$, 
  $\tan \beta = 10$, $\mathrm{sign}(\mu) = (-)$,
   $n_{L_i} = -3$, $n_{E_i} = -1$,$n_{Q_i} = 0$, $n_{U_i} = -2$, $n_{D_i} = 1$
and $n_{H_1} =n_{H_2}=-1$. 
[The widths of the bands indicate the $1 \, \sigma$ CL.]
}}
\label{fig:String}
\end{figure*} 

In the case of the gaugino mass parameters 
the next--to--leading order effects
induce a splitting proportional to the beta functions $b_i$ which is large
enough to be ``measured'' at future collider experiments as demonstrated
in \fig{fig:String}(a) and (b).
In the 
case of the scalar mass parameters  the next--to--leading
order contributions generate small departures from non-universality
between the generations even if the corresponding modular weight is
generation independent. These departures are proportional to the Yukawa
couplings squared so that the third generation, in particular
the stop sector, is mainly affected.

Scenarios have been found in which the phenomenological unification
of the three gauge couplings can be reconciled with a string mass scale
which is an order of magnitude larger than the unification scale 
\cite{Ibanez:1991zv}:
\begin{eqnarray}
\alpha^{-1}_i(M_U) &=& \alpha^{-1}(M_{String}) + \Delta\alpha^{-1}_i
\end{eqnarray}
The corrections $\Delta\alpha^{-1}_i$ depend on the value of the moduli fields
and the modular weights:
\begin{eqnarray}
\Delta\alpha^{-1}_i = \frac{1}{4 \pi} (b_i' - b_{GS}) \log|\eta(t)|^4
\end{eqnarray}
where $\eta(t)$ is the Dedekind $\eta$ function and 
\begin{eqnarray}
b_3' &=& 9 + \sum_{i=1}^3 \left( 2 n_{Q_i} + n_{U_i} + n_{D_i} \right)\\
b_2' &=& 15 + \sum_{i=1}^3 \left( 3 n_{Q_i} + n_{L_i} \right)
            + n_{H_1} + n_{H_2}\\
b_1' &=& \frac{99}{5} + \frac{1}{5}\
      \sum_{i=1}^3 \left( n_{Q_i} + 8 n_{U_i} + 2 n_{D_i} + 3 n_{L_i}
                       + 6  n_{E_i} \right)
            + \frac{3}{5} \left(n_{H_1} + n_{H_2}\right)
\end{eqnarray}
as compared to the one--loop $SU(3)\times SU(2) \times U(1)$ $\beta$--functions
$(b_3, b_2, b_1)$ $=$ $33/5, 1, -3$.

We have analyzed  a mixed dilaton/moduli
superstring scenario but with dominating dilaton
field component, $\sin^2 \theta = 0.9$, and with different couplings of the
moduli field to the (L,R) sleptons, the (L,R) squarks and to the Higgs
fields, corresponding to O--I representation $n_{L_i} = -3$, $n_{E_i} = -1$, 
$n_{H_1} =n_{H_2}=-1$,
 $n_{Q_i} = 0$, $n_{D_i} = 1$ and $n_{U_i} = -2$, that is one out of several
assignments that is adopted quite frequently in the literature.
The gravitino mass is chosen to be 180~GeV in this analysis.
We find for the low energy data: BR($b\to s \gamma) = 3.1 \cdot 10^{-4}$, 
$\Delta[g - 2]_\mu = 14 \cdot 10^{-10}$, $\Delta \rho = 13 \cdot 10^{-5}$;
and $\Omega h^2=0.25$.

Given this set of superstring induced parameters, the evolution of the
gaugino and scalar mass parameters is displayed in \fig{fig:String}.
The pattern of the trajectories is remarkably different from other
scenarios.  The breaking of universality in the gaugino
sector, induced by string threshold corrections, is pronounced, 
cf.~Table~\ref{tab:parvalues_string}. 

\begin{table}
\caption[]{\it Representative gaugino/scalar mass parameters and couplings
as determined
at the electroweak
scale and evolved to the GUT scale in the string scenario;
based on LHC
and LC simulations. $M^2_{L_{1,3}}$, $M^2_{Q_{1,3}}$ are the slepton and
squark isodoublet parameters of
the first and third family whereas $M^2_{E_{1,3}}$, $M^2_{U_{1,3}}$ and
$M^2_{D_{1,3}}$ are the  the slepton and
squark isosinglet parameters of
the first and third family. [The errors quoted correspond to 1$\sigma$.]}
\label{tab:parvalues_string}
\begin{center}
\begin{tabular}{c||c|c}
 & exp.~input &  GUT value \\ \hline  \hline
 $M_1$~[GeV] & -124.98 $\pm$ 0.29 &  -303.22 $\pm$ 0.65 \\
 $M_2 $~[GeV] & -231.00 $\pm$ 0.50      &  -299.64$\pm$ 0.52 \\
 $M_3 $~[GeV] & -677.3 $\pm$  7.6   &  -292.4   $\pm$  3.3 \\ \hline
$\mu$         &   -377.59 $\pm$ 0.29           & -375.5$\pm$ 1.2   \\
\hline
 $M^2_{L_1} $~[GeV$^2$] & ( 6.354 $\pm$ $0.005)\cdot 10^4$
                &  (2.17 $\pm$ $0.43) \cdot 10^4 $ \\
 $M^2_{E_1} $~[GeV$^2$] & (3.739 $\pm$ $0.005) \cdot 10^4$
  &  (2.88 $\pm 0.86)  \cdot 10^4 $ \\
\hline
 $M^2_{Q_1} $~[GeV$^2$] &  (4.16$\pm$ $0.09)\cdot 10^5$
               &  (3.1 $\pm 1.3) \cdot 10^4$ \\
 $M^2_{U_1} $~[GeV$^2$] &  (3.80$\pm 0.12)\cdot 10^5$
               &  (2.5 $\pm 1.9) \cdot 10^4$ \\
 $M^2_{D_1} $~[GeV$^2$] &  (3.88$\pm 0.13)\cdot 10^5$
               &  (3.5 $\pm 1.7)  \cdot 10^4$
\\ \hline
 $M^2_{L_3} $~[GeV$^2$] & (5.635$\pm 0.039)\cdot 10^4$
   &  (2.18 $\pm 0.46)  \cdot 10^4 $  \\
 $M^2_{E_3} $~[GeV$^2$] & (2.253$\pm 0.024)\cdot 10^4$
  &  (2.90 $\pm 0.93) \cdot 10^4 $  \\
\hline
 $M^2_{Q_3} $~[GeV$^2$] &  (3.28$\pm 0.13) \cdot 10^5$
  &  (3.2 $\pm$ $2.1)  \cdot 10^4 $ \\
 $M^2_{U_3} $~[GeV$^2$] &  (2.58$\pm 0.15)\cdot 10^5$
  &  (2.6 $\pm 3.3)  \cdot 10^4 $ \\
 $M^2_{D_3} $~[GeV$^2$] &  (3.53$\pm 0.15)\cdot 10^5$
  &  (3.5 $\pm 1.8)  \cdot 10^4 $ \\
\hline
 $M^2_{H_1} $~[GeV$^2$] &  (3.80$\pm 0.82)\cdot 10^3$  &
  (2.85 $\pm 0.62)  \cdot 10^4 $ \\
 $M^2_{H_2}$~[GeV$^2$] &  (-1.429$\pm0.004)\cdot 10^5$ &
  (3.1 $\pm 2.7) \cdot 10^4 $\\
 $A_t $~[GeV] & 452 $\pm$ 17   &  -96 $\pm$ 64   \\ \hline
 $\tan\beta$ & 9.93 $\pm$ 0.88 & --- \\ \hline
\end{tabular}
\end{center}
\end{table}

In fact, the differences can be exploited to determine superstring parameters
as argued above.
The number of observables in the set of gauge couplings $g_a$, gaugino
masses $M_a$ and scalar masses $M_{\tilde j}$ exceeds the number of
parameters in the superstring effective field theory: the gravitino
mass $m_{3/2}$, the dilaton/moduli mixing angle $\sin \theta$, the 
ground--state value of the moduli field $\langle T \rangle$ and 
the ground--state 
value of the dilaton field $\langle S \rangle$. The latter  is at tree--level
directly related to the string coupling: $1/g_s^2 = \langle S \rangle$.  

\begin{table}
\caption[]{\it Comparison of the experimentally reconstructed values with the
               ideal fundamental parameters in a specific example for a string 
               effective field theory.} 
\label{tab:parameters_string2}
\begin{center}
\begin{tabular}{c||c|rcl}
Parameter           & Ideal & \multicolumn{3}{c}{Reconstructed} \\ \hline\hline
$m_{3/2}$           &  180  &     179.9 & $\pm$ & 0.4 \\
$\langle S \rangle$ &   2   &      1.998 & $\pm$ & 0.006 \\
$\langle T \rangle$ &  14   &      14.6 & $\pm$ & 0.2 \\
$\sin^2\theta$        & 0.9 &      0.899 & $\pm$ & 0.002 \\
$g_s^2$             & 0.5   &      0.501 & $\pm$ & 0.002 \\
$\delta_{GS}$       &   0   &      0.1 & $\pm$ & 0.4 \\ \hline
$n_L$               &  -3   &      -2.94 & $\pm$ & 0.04 \\
$n_E$               &  -1   &     -1.00 & $\pm$ & 0.05 \\
$n_Q$               &   0   &     0.02 & $\pm$ & 0.02 \\
$n_U$               &  -2   &     -2.01 & $\pm$ & 0.02 \\
$n_D$               &  +1   &      0.80 & $\pm$ & 0.04 \\
$n_{H_1}$           &  -1   &      -0.96 & $\pm$ & 0.06 \\
$n_{H_2}$           &  -1   &      -1.00 & $\pm$ & 0.02 \\ \hline
$\tan \beta$        &  10   &      10.00 & $\pm$ & 0.13 \\ 
\end{tabular}
\end{center}
\end{table}

Based on the ``experimental'' input observables, the fundamental parameters 
of the string effective field theory can be reconstructed; the reconstructed 
values are compared with the ideal values in 
Table~\ref{tab:parameters_string2}.
The errors for  the basic parameters  $\sin \theta$, 
$\langle T \rangle/m_{3/2}$ are displayed in Figs.~\ref{fig:String}(b).

Thus, high-precision measurements at high energy proton and $e^+e^-$ 
linear colliders provide access to crucial derivative parameters 
in string theories.

\section{Conclusions}

In supersymmetric theories stable extrapolations can be 
performed from the electroweak scale to the Grand Unification 
scale close to the Planck scale. This feature has been 
compellingly demonstrated in the evolution of the three 
gauge couplings to the unification point in the minimal
supersymmetric theory.

Such extrapolations are made possible by high-precision 
measurements of the low-energy parameters. The operation 
of the $e^+e^-$ colliders LEP and SLC has been crucial in this
context. In the near future an enormous extension of this 
area will be possible if measurements at LHC and prospective
$e^+e^-$ linear colliders are combined to draw, if realized in Nature,
a comprehensive high-precision picture of the supersymmetric 
particles and their interactions. Based merely on measurements
at low energies, the parameters of the theory can be evolved
to high scales by means of renormalization group techniques.

Supersymmetric theories and their breaking mechanisms have the
simplest structures and the greatest regularities at high scales. 
Extrapolations to high scales are therefore crucial to uncover 
the regularities. The bottom-up approach in the extrapolation 
of parameters measured at low scales to the high scales provides
the most transparent picture. In this way the basis 
of the SUSY breaking 
mechanism can be explored and the crucial elements of the fundamental 
supersymmetric theory can be reconstructed. The method can thus 
be used to explore particle physics phenomena at a scale where, 
eventually, particle physics is linked to gravity.

Apart from other examples, we have focused on two interesting
scenarios in this approach. The universality of gaugino and scalar
mass parameters in minimal supergravity can be demonstrated very 
clearly if realized in the supersymmetric theory. Small deviations 
from universality, on the other hand, may be exploited to measure 
the fundamental parameters in superstring effective field theories, 
i.e. the strength of dilaton and moduli fields, their mixing and
the modular weights. In this way, high-precision extrapolations of 
gauge and supersymmetric parameters can establish direct contact 
between superstring theory and experiment. 

Many more refinements of the theoretical calculations and future
experimental analyses will be necessary to expand the picture we
have drawn in this first attempt. However, the prospect of exploring
elements of the ultimate unification of the interactions provides
a strong impetus to this direction.

\section*{Acknowledgements}

We are very grateful for enlightening discussions with J.~Kalinowski, 
G.L.~Kane, B.D.~Nelson and H.P.~Nilles. 
G.A.B. is grateful for the hospitality extended to him in long-term
visits to CERN and DESY, for support from the British Council ARC 
programme and Intas grant 00-00679.
W.P. is supported by the Erwin
Schr\"odinger fellowship No. J2095 of the `Fonds zur
F\"orderung der wissenschaftlichen Forschung' of Austria FWF and
partly by the Swiss `Nationalfonds'.

\begin{appendix}

\section{One--loop RGEs}

In this first appendix 
we collect the one--loop renormalization group equations (RGEs) including 
right-handed neutrinos.

Using the notation for the gauge and Yukawa couplings
\begin{eqnarray}
\label{eq:alpha}
\alpha_i = \frac{g_i^2}{16\pi^2}, \ i=1,2,3; \ \  \ Y_k =
\frac{y_k^2}{16\pi^2}, \ k=t,b,\tau,\nu
\end{eqnarray}
the one-loop RG equations can be written as 
\begin{eqnarray}
\dot{\alpha}_i &=& -b_i\alpha_i^2, \label{a}\\ \dot{Y}_k &=&
Y_k(\sum_{i}c_{ki}\alpha_i - \sum_{l}a_{kl}Y_l), \label{y}
\end{eqnarray}
where  the dot denotes the derivative with respect to $t= \log M_{GUT}^2/Q^2$,
 and
\begin{eqnarray}
b_i&=&\{33/5,1,-3 \}, \\
c_{ti}&=& \{13/15,3,16/3 \}, \ \
c_{bi} = \{7/15,3,1 6/3 \},  \\ 
c_{\tau i}&=&\{9/5,3,0 \}, \ \
c_{\nu i} = \{3/5,3,0 \}, \\
a_{tl}&=&\{6,1,0,1 \}, \ \
a_{bl} = \{1,6,1,0 \}, \\ 
a_{\tau l}&=&\{0,3,4,1\}, \ \ 
a_{\nu l} = \{3,0,1,4\} \, ,
\end{eqnarray}
while the RGEs for the gaugino mass parameters and the $A$-parameters read
\begin{eqnarray}
\dot{M}_i &=& - b_i \alpha_i M_i \, ,\\ 
\dot{A}_k &=& \sum_{i}c_{ki}\alpha_i M_i - \sum_{l}a_{kl}A_l.
\end{eqnarray}
The RGEs for the soft SUSY breaking mass parameters of the third generation
and the Higgs mass parameters are given by:
\begin{eqnarray}
\label{eq:RGEMsq}
\dot{M}_{L_3} &=& - 2 Y_\tau X_\tau - 2 Y_\nu X_\nu
 + \frac{6}{5} \alpha_1 M^2_1 + 6 \alpha_2 M^2_2 + \frac{3}{5} S
\, ,  \\
\dot{M}_{\nu_{R3}} &=& - 4 Y_\nu X_\nu \, ,\\
\dot{M}_{E_3} &=& - 4 Y_\tau X_\tau
               + \frac{24}{5} \alpha_1 M^2_1 - \frac{6}{5} S \, , \\
\dot{M}_{Q_3} &=& - 2 Y_b X_b - 2 Y_t X_t 
+ \frac{2}{15} \alpha_1 M^2_1 + 6 \alpha_2 M^2_2
 + \frac{16}{3} \alpha_3 M^2_3 - \frac{1}{5} S
 \, , \\
\dot{M}_{U_3} &=& - 4 Y_t X_t
 + \frac{32}{15} \alpha_1 M^2_1 + \frac{16}{3} \alpha_3 M^2_3 + \frac{4}{5} S
 \, , \\
\dot{M}_{D_3} &=& - 4 Y_b X_b
 + \frac{8}{15} \alpha_1 M^2_1 + \frac{16}{3} \alpha_3 M^2_3 - \frac{2}{5} S
 \, , \\
\dot{M}_{H_1} &=& - 6 Y_b X_b - 2 Y_\tau X_\tau
 + \frac{6}{5} \alpha_1 M^2_1 + 6 \alpha_2 M^2_2 + \frac{3}{5} S \, ,
  \\
\dot{M}_{H_2} &=& - 6 Y_t X_t - 2 Y_\nu X_\nu
 + \frac{6}{5} \alpha_1 M^2_1 + 6 \alpha_2 M^2_2 - \frac{3}{5} S
 \, ,
\end{eqnarray}
with
\begin{eqnarray}
X_t&=& M^2_{Q_3}+M^2_{U_3}+M^2_{H_2}+A_t^2 \, ,\\
X_b&=& M^2_{Q_3}+m_D^2+M^2_{H_1}+A_b^2 \, ,\\
X_\tau &=& M^2_{L_3}+M^2_{E_3}+M^2_{H_1}+A_\tau^2 \, ,\\
X_\nu &=& M^2_{L_3}+M^2_{\nu _{R3}} +M^2_{H_2}+A_\nu^2 \, ,\\
\label{eq:S}
S &=& M^2_{H_2} -  M^2_{H_2}
    +  \sum_{i=1}^3  \bigg( M^2_{Q_i} - M^2_{L_i}
  - 2 M^2_{U_i} + M^2_{D_i} + M^2_{E_i} \bigg) \, .
\end{eqnarray}
The evolution equations for the first two generations are obtained by
replacing appropriately the corresponding parameters and Yukawa couplings.

\section{Solutions of the one--loop RGEs}
\label{sec:AppRGE}

In the following subsections 
we present the analytical solutions to the 1-loop RGEs
including Yukawa couplings using the procedure of 
Ref.~\cite{Kazakov:2000pe}. We also include the generic trace term 
$S$ (see \eq{eq:S}) in the solutions which had been neglected in 
Ref.~\cite{Kazakov:2000pe}.
In this appendix we mark all quantities defined at the the GUT-scale
 $M_{GUT}$ with a subscript G.

\subsection{mSUGRA Boundary Conditions at the GUT-scale}
\label{sec:AppRGEGUT}

The solutions for the case of the MSSM are summarized first for proper
reference.
The solution for the gauge couplings and Yukawa couplings are given by:
\begin{eqnarray}
\label{eq:alphaGUT}
 \alpha_{i}(t) &=&  \frac{\alpha_{i,G}}{1 + b_i \alpha_{i,G} t} \\
\label{eq:yukawaGUT}
 Y_k(t) &=& \frac{Y_{k,G} u_k}{1+a_{kk} Y_{k,G} \int_0^t u_k},
\end{eqnarray}
where the functions $ u_k$ obey the integral system of equations
\begin{eqnarray}
u_t &=& \frac{E_t}{(1+6Y_{b,G}\int_0^t u_b)^{1/6}} \, , \\
u_b &=& \frac{E_b}
       {(1+6Y_{t,G}\int_0^t u_t)^{1/6} (1+4 Y_{\tau,G} \int_0^t u_\tau)^{1/4}}
        \, , \\ 
u_\tau &=&\frac{E_\tau}{(1+6Y_{b,G}\int_0^t u_b)^{1/2}} \, ,
\end{eqnarray}
and the functions $E_k$ denote the products
\begin{equation}
E_k= \prod_{i=1}^3(1+b_i \alpha_{i,G} t)^{c_{ki}/b_i} \, . \label{e}
\end{equation}
The system of integral equations can be solved iteratively and
a discussion on the convergence can be found in Ref.~\cite{Kazakov:2000pe}.

The gaugino mass parameters and the $A_k$ parameters are given by
\begin{eqnarray}
 M_{i}(t) &=&  \frac{M_{i,G}}{1 + b_i \alpha_{i,G} t}
            = \frac{\alpha_i(t)}{\alpha_{i,G}}M_{i,G} \, , \\
A_k &=& -e_k + \frac{A_{k,G}/Y_{k,G} +a_{kk}\int u_ke_k}
                    {1/Y_{k,G} +a_{kk}\int u_k} \, ,
\end{eqnarray}
with the coefficients
\begin{eqnarray}
 e_t &=& \tilde{F}_t
    + \frac{A_{b,G}\int u_b-\int u_be_b}{ 1/Y_{b,G}+6\int u_b } \, ,
\\
 e_b &=&  \tilde{F}_b 
 +\frac{A_{t,G}\int u_t-\int u_te_t}{ 1/Y_{t,G}+6\int u_t}
 + \frac{A_{\tau,G}\int u_\tau-\int u_\tau e_\tau}{1/Y_{\tau,G} +4\int u_\tau}
    \, ,
\\
 e_\tau &=& \tilde{F}_\tau
  + 3\frac{A_{b,G}\int u_b -\int u_be_b}{1/Y_{b,G}+6\int u_b} \, ,
\\
\tilde{F}_k &=& t \sum^3_{i=1} c_{k i} M_{i,G} \alpha_i(t) \, . 
\end{eqnarray}
The mass parameters of the first two generations ($k=1,2$) can be expressed as
\begin{eqnarray}
 M^2_{L_k}(t)  &=&   M^2_{L_k,G}
   +  \frac{3}{2} f_{2}(t) + \frac{3}{10} f_{1}(t)
   + \frac{3}{5} S'(t) \\
 M^2_{E_k}(t) &=&  M^2_{E_k,G}
        +  \frac{6}{5} f_{1}(t) - \frac{6}{5} S'(t) \\
 M^2_{Q_k}(t) &=& M^2_{Q_k,G}
   + \frac{8}{3} f_{3}(t) + \frac{3}{2} f_{2}(t)
       + \frac{1}{30} f_{1}(t) - \frac{1}{5} S'(t)  \\
 M^2_{U_k}(t) &=& M^2_{U_k,G} 
   +\frac{8}{3} f_{3}(t) + \frac{8}{15} f_{1}(t)
   + \frac{4}{5} S'(t) \\
 M^2_{D_k}(t) &=& M^2_{D_k,G} 
   + \frac{8}{3} f_{3}(t) + \frac{2}{15} f_{1}(t)
  - \frac{2}{5} S'(t) 
\end{eqnarray}
with
\begin{eqnarray}
  f_i(t) &=&  \frac{M_{i,G}^2}{b_i}
       \left(1 - \frac{1}{(1 + \alpha_{i,G} b_i t)^2} \right) \\
S'(t) &=& \frac{1}{2 b_1} \left( S(t) - S(M_U) \right) \\
S(t) &=& S(M_U) (1 + \beta_1 t)^2 \\
S(M_U) &=& M^2_{H_1,G} -  M^2_{H_2,G}
    +  \sum_{i=1}^3  \bigg( M^2_{Q_i,G} - M^2_{L_i,G}
  - 2 M^2_{U_i,G} + M^2_{D_i,G} + M^2_{E_i,G} \bigg) \, ,
\end{eqnarray}
in agreement with Ref.~\cite{RGE}. The mass 
parameters for the third generation and the Higgs mass parameters are
involved owing to the Yukawa couplings:
\begin{eqnarray}
M^2_{L_3} &=& M^2_{L_3,G}
        +\frac{80f_3+123f_2-103/5f_1}{122} - \frac{3}{5} S'(t) 
   + \frac{3 \Delta \Sigma_t - 18  \Delta \Sigma_b +35 \Delta \Sigma_\tau }
            {122} \, , \\
M^2_{E_3} &=& M^2_{E_3,G}
       + \frac{80f_3-60f_2+16f_1}{61} + \frac{6}{5} S'(t)
    + \frac{3 \Delta \Sigma_t - 18 \Delta \Sigma_b + 35 \Delta \Sigma_\tau}
             {61} \, , \\
M^2_{Q_3} &=& M^2_{Q_3,G}
   + \frac{128 f_3 + 87f_2 - 11f_1}{122} + \frac{1}{5} S'(t)
   + \frac{17\Delta \Sigma_t + 20 \Delta \Sigma_b -5 \Delta \Sigma_\tau }
            {122}  \, ,\\
M^2_{U_3} &=& M^2_{U_3,G}
          +\frac{72f_3-54f_2+72/5f_1}{61} - \frac{4}{5} S'(t) 
     + \frac{21 \Delta \Sigma_t-4 \Delta \Sigma_b + \Delta \Sigma_\tau}
              {61} \, ,\\
M^2_{D_3} &=& M^2_{D_3,G} 
         +\frac{56 f_3-42f_2+56/5f_1}{61} + \frac{2}{5} S'(t)
   + \frac{-4 \Delta \Sigma_t+24 \Delta \Sigma_b -6 \Delta \Sigma_\tau }
            {61} \, ,\\
M^2_{H_1} &=& M^2_{H_1,G}
        +\frac{-240f_3-3f_2-57/5f_1}{122} - \frac{3}{5} S'(t)
  +   \frac{-9 \Delta \Sigma_t+54 \Delta \Sigma_b +17 \Delta \Sigma_\tau }
             {122} \, , \\
M^2_{H_2} &=& M^2_{H_2,G}
      +\frac{-272f_3+21f_2-89/5f_1}{122} + \frac{3}{5} S'(t)
  + \frac{63 \Delta \Sigma_t-12 \Delta \Sigma_b +3 \Delta \Sigma_\tau }
           {122} \, ,
\end{eqnarray}
with
\begin{eqnarray}
\Delta \Sigma_k &=& \Sigma_k(t) - \Sigma_{k,G} \, ,\\
\Sigma_t &=& M_{Q_3}^2 + M_{U_3}^2  + M^2_{H_ 2} \, , \\ 
\Sigma_b &=& M_{Q_3}^2 + M_{D_3}^2  + M^2_{H1} \, , \\ 
\Sigma_\tau &=& M_{L_3}^2 + M_{E_3}^2 + M^2_{H1} \, .
\end{eqnarray}
The explicit solution for $\Sigma_k$ read as:
\begin{eqnarray}
 \Sigma_k &=& \xi_k + A_k^2 + 2 e_k A_k 
   - \frac{A_{k,G}^2 / Y_{k,G} -\Sigma_{k,G} / Y_{k,G} + a_{kk}\int u_k\xi_k}
          {1/Y_{k,G}+a_{kk}\int u_k} \, ,
\end{eqnarray}
with
\begin{eqnarray}
 \xi_t &=& \tilde{E}_t
        + 2 \tilde{F}_t
          \frac{A_{b,G}\int u_b-\int u_be_b}{1/Y_{b,G}+6\int u_b} 
   + 7\left(\frac{A_{b,G}\int u_b - \int u_b e_b}{ 1/Y_{b,G} +6\int u_b}
                 \right)^2  \nonumber \\
   &-& \frac{(\Sigma_{b,G}+A_{b,G}^2)\int u_b -2 A_{b,G}\int u_be_b
             +\int u_b\xi_b} {1/Y_{b,G} +6\int u_b },  \\
 \xi_b &=& \tilde{E}_b 
    + 2 \tilde{F}_b
          \left[ \frac{A_{t,G} \int u_t-\int u_te_t}{1/Y_{t,G} +6\int u_t}
               + \frac{A_{\tau,G} \int u_\tau-\int u_\tau e_\tau}
                      { 1/Y_{\tau,G} +4\int u_\tau}
          \right] \nonumber \\
    &+& 7 \left(\frac{A_{t,G}\int u_t-\int u_te_t}{1/Y_{t,G} +6\int u_t}
            \right)^2
    + 5 \left(\frac{A_{\tau,G}\int u_\tau-\int u_\tau e_\tau}
                     {1/Y_{\tau,G} +4\int u_\tau} \right)^2
          \nonumber \\
    &+& 2\left(\frac{A_{t,G} \int u_t-\int u_te_t}{1/Y_{t,G} +6\int u_t}\right)
          \left(\frac{A_{\tau,G} \int u_\tau-\int u_\tau e_\tau}
                     { 1/Y_{\tau,G} +4\int u_\tau}\right) 
    - \frac{(\Sigma_{t,G} + A_{t,G}^2)\int u_t -2 A_{t,G} \int u_te_t
            +\int u_t\xi_t}
           { 1/Y_{t,G} +6\int u_t } \nonumber \\
    &-& \frac{ (\Sigma_{\tau,G}+A_{\tau,G}^2) \int u_\tau
            - 2A_{\tau,G}\int u_\tau e_\tau +\int u_\tau\xi_\tau }
          {1/Y_{\tau,G} +4\int u_\tau } \, ,  \\
 \xi_\tau &=& \tilde{E}_\tau
       + 6 \tilde{F}_\tau 
         \frac{A_{b,G}\int u_b-\int u_be_b}{1/Y_{b,G}+6\int u_b} 
 + 27\left(\frac{A_{b,G}\int u_b-\int u_be_b}{1/Y_{b,G}+6\int u_b} \right)^2 
       \nonumber\\
 &-& 3 \frac{ (\Sigma_{b,G}+A_{b,G}^2)\int u_b-2A_{b,G}\int u_b e_b
                  +\int u_b\xi_b }{ 1/Y_{b,G} + 6\int u_b } \, \\
\tilde{E}_k &=& t^2 \left(\sum_{i=1}^3 c_{ki} \alpha_i M_{i,G} \right)^2
             + 2t \sum_{i=1}^3 c_{ki} \alpha_i M_{i,G}^2
             - t^2 \sum_{i=1}^3 c_{ki} b_i \alpha_i^2 M_{i,G}^2 \, .
\end{eqnarray}
Finally we express $t_Z = \log(M_{GUT}^2/m^2_Z)$ and $\alpha_G$ in terms of
observables at the electroweak scale, using \eq{eq:alphaGUT}, by
\begin{eqnarray}
t_Z = \frac{4 \pi}{(b_1-b_2)\alpha(m_Z)}
     \left( \frac{3 \cos^2 \vartheta_W}{5} - \sin^2 \vartheta_W \right)
\end{eqnarray}
and similarly for the gauge coupling at the GUT scale:
\begin{eqnarray}
\alpha_G = \frac{5\alpha(m_Z)}{3}
      \frac{b_1-b_2}{\frac{5}{3}b_1 \sin^2 \Theta_W - b_2 \cos^2 \Theta_W}
\end{eqnarray}

\subsection{Universal SUGRA Boundary Conditions at the GUT-scale 
            including Right-handed Neutrinos}

Those formulae are noted in this subsection which are changed compared to the
previous section in the range between $M_U$ and $M_{\nu_R}$. 
Below $M_{\nu_R}$, these quantities have the same form as given above. 
In addition
we note also the equations related to the right-handed neutrinos,
\begin{eqnarray}
u_t &=& \frac{E_t}
       {(1+6Y_{b,G}\int_0^t u_b)^{1/6} (1+4 Y_{\nu,G} \int_0^t u_\nu)^{1/4}}
        \, , \\
u_\tau &=&\frac{E_\tau}
     {(1+6Y_{b,G}\int_0^t u_b)^{1/2} (1+4 Y_{\nu,G} \int_0^t u_\nu)^{1/4}}
         \, , \\
u_\nu &=&\frac{E_\nu}
     {(1+6Y_{t,G}\int_0^t u_t)^{1/2} (1+4 Y_{\tau,G} \int_0^t u_\tau)^{1/4}}
         \, , 
\end{eqnarray}
\begin{eqnarray}
 e_t &=& \tilde{F}_t
    + \frac{A_{b,G}\int u_b-\int u_b e_b}{1/Y_{b,G}+6\int u_b } 
    + \frac{A_{\nu,G}\int u_\nu-\int u_\nu e_\nu}{1/Y_{\nu,G}+4\int u_\nu } 
\, , 
\\
 e_\tau &=& \tilde{F}_\tau
  + 3\frac{A_{b,G}\int u_b -\int u_b e_b}{1/Y_{b,G}+6\int u_b}
    + \frac{A_{\nu,G}\int u_\nu-\int u_\nu e_\nu}{1/Y_{\nu,G}+4\int u_\nu } 
\, , 
\\
 e_\nu &=& \tilde{F}_\nu
  + 3\frac{A_{t,G}\int u_t -\int u_t e_t}{1/Y_{t,G}+6\int u_t}
  + \frac{A_{\tau,G}\int u_\tau-\int u_\tau e_\tau}{1/Y_{\tau,G}+4\int u_\tau}
\, . 
\end{eqnarray}
\begin{eqnarray}
 \xi_t &=& \tilde{E}_t 
      + 2 \tilde{F}_t
      \left( \frac{A_{b,G}\int u_b-\int u_be_b}{1/Y_{b,G}+6\int u_b}
       +  \frac{A_{\nu,G}\int u_\nu-\int u_\nu e_\nu}{1/Y_{\nu,G}+4\int u_\nu}
      \right)
   + 7\left(\frac{A_{b,G}\int u_b - \int u_b e_b}{ 1/Y_{b,G} +6\int u_b}
                 \right)^2   \nonumber \\
    &+& 5 \left(\frac{A_{\nu,G}\int u_\nu-\int u_\nu e_\nu}
                     {1/Y_{\nu,G} +4\int u_\nu} \right)^2
   + 2\left(\frac{A_{b,G} \int u_b-\int u_b e_b}{1/Y_{b,G} +6\int u_b}\right)
          \left(\frac{A_{\nu,G} \int u_\nu-\int u_\nu e_\nu}
                     { 1/Y_{\nu,G} +4\int u_\nu}\right) \nonumber \\
   &-& \frac{(\Sigma_{b,G}+A_{b,G}^2)\int u_b -2 A_{b,G}\int u_be_b
             +\int u_b\xi_b} {1/Y_{b,G} +6\int u_b }, \nonumber \\
    &-& \frac{ (\Sigma_{\nu,G}+A_{\nu,G}^2) \int u_\nu
            - 2A_{\nu,G}\int u_\nu e_\nu +\int u_\nu\xi_\nu }
          {1/Y_{\nu,G} +4\int u_\nu } \, ,  \\
 \xi_\tau &=& \tilde{E}_\tau \nonumber 
      + 2 \tilde{F}_\tau
      \left(3 \frac{A_{b,G}\int u_b-\int u_be_b}{1/Y_{b,G}+6\int u_b}
       +  \frac{A_{\nu,G}\int u_\nu-\int u_\nu e_\nu}{1/Y_{\nu,G}+4\int u_\nu}
      \right) \nonumber \\
 &+& 27\left(\frac{A_{b,G}\int u_b-\int u_be_b}{1/Y_{b,G}+6\int u_b} \right)^2 
    + 5 \left(\frac{A_{\nu,G}\int u_\nu-\int u_\nu e_\nu}
                     {1/Y_{\nu,G} +4\int u_\nu} \right)^2
          \nonumber \\
   &+& 6\left(\frac{A_{b,G} \int u_b-\int u_b e_b}{1/Y_{b,G} +6\int u_b}\right)
          \left(\frac{A_{\nu,G} \int u_\nu-\int u_\nu e_\nu}
                     { 1/Y_{\nu,G} +4\int u_\nu}\right) \nonumber \\
 &-& 3 \frac{ (\Sigma_{b,G}+A_{b,G}^2)\int u_b-2A_{b,G}\int u_b e_b
                  +\int u_b\xi_b }{ 1/Y_{b,G} + 6\int u_b } \, \nonumber \\
    &-& \frac{ (\Sigma_{\nu,G}+A_{\nu,G}^2) \int u_\nu
            - 2A_{\nu,G}\int u_\nu e_\nu +\int u_\nu\xi_\nu }
          {1/Y_{\nu,G} +4\int u_\nu } \, ,  \\
 \xi_\nu &=& \tilde{E}_\nu 
      + 2 \tilde{F}_\nu
  \left(3 \frac{A_{t,G}\int u_t-\int u_t e_t}{1/Y_{t,G}+6\int u_t}
   + \frac{A_{\tau,G}\int u_\tau-\int u_\tau e_\tau}{1/Y_{\tau,G}+4\int u_\tau}
  \right) \nonumber \\
 &+& 27\left(\frac{A_{t,G}\int u_t-\int u_te_t}{1/Y_{t,G}+6\int u_t} \right)^2 
    + 5 \left(\frac{A_{\tau,G}\int u_\tau-\int u_\tau e_\tau}
                     {1/Y_{\tau,G} +4\int u_\tau} \right)^2
          \nonumber \\
   &+& 6\left(\frac{A_{t,G} \int u_t-\int u_t e_t}{1/Y_{t,G} +6\int u_t}\right)
          \left(\frac{A_{\tau,G} \int u_\tau-\int u_\tau e_\tau}
                     { 1/Y_{\tau,G} +4\int u_\tau}\right) \nonumber \\
 &-& 3 \frac{ (\Sigma_{t,G}+A_{t,G}^2)\int u_t-2A_{t,G}\int u_t e_t
                  +\int u_t\xi_t }{ 1/Y_{t,G} + 6\int u_t } \, \nonumber \\
    &-& \frac{ (\Sigma_{\tau,G}+A_{\tau,G}^2) \int u_\tau
            - 2A_{\tau,G}\int u_\tau e_\tau +\int u_\tau\xi_\tau }
          {1/Y_{\tau,G} +4\int u_\tau } \, .
\end{eqnarray}
\begin{eqnarray}
\Sigma_\nu &=& M_{L_3}^2 + M_{\nu_{3,R}}^2 + M^2_{H_2} \, .
\end{eqnarray}

\end{appendix}

\end{document}